\documentclass[twocolumn,tighten,times,trackchanges]{aastex63}
\usepackage{times, graphicx, subfigure, latexsym, amssymb, amsmath, fancyhdr, epsf, upgreek}
\usepackage{natbib}
\usepackage{multirow}
\usepackage{color}
\usepackage{rotating}
\usepackage{subfigure}
\usepackage{cprotect}
\usepackage{afterpage}
\usepackage{xspace}
\usepackage{longtable}
\newcolumntype{L}{>{\centering\arraybackslash}m{4cm}}
\newcommand{\psrchive}{\textsc{psrchive}\xspace}
\newcommand{\pypulse}{\textsc{pypulse}\xspace}
\newcommand{\psrsigsim}{\textsc{psrsigsim}\xspace}

\newcommand{\pint}{\textsc{pint}\xspace}
\newcommand{\psrfits}{\textsc{psrfits}\xspace}
\newcommand{\DM}{\rm{DM}\xspace}
\newcommand{\DMX}{\rm{DMX}\xspace}
\newcommand{\FD}{\rm{FD}\xspace}

\shorttitle{Frequency-Dependent Effects with the PsrSigSim}
\shortauthors{B. J. Shapiro-Albert et al.}

\begin{document}

\title{A Study in Frequency-Dependent Effects on Precision Pulsar Timing Parameters with the Pulsar Signal Simulator}

\correspondingauthor{Brent J. Shapiro-Albert}
\email{bjs0024@mix.wvu.edu}

\author[0000-0002-7283-1124]{B.~J.~Shapiro-Albert}
\affiliation{Department of Physics and Astronomy, West Virginia University, P.O. Box 6315, Morgantown, WV 26506, USA}
\affiliation{Center for Gravitational Waves and Cosmology, West Virginia University, Chestnut Ridge Research Building, Morgantown, WV 26505, USA}

\author[0000-0003-2742-3321]{J.~S.~Hazboun}
\affiliation{Physical Sciences Division, University of Washington Bothell, 18115 Campus Way NE, Bothell, WA 98011, USA}

\author[0000-0001-7697-7422]{M.~A.~McLaughlin}
\affiliation{Department of Physics and Astronomy, West Virginia University, P.O. Box 6315, Morgantown, WV 26506, USA}
\affiliation{Center for Gravitational Waves and Cosmology, West Virginia University, Chestnut Ridge Research Building, Morgantown, WV 26505, USA}

\author[0000-0003-0721-651X]{M. T. Lam}
\affiliation{School of Physics and Astronomy, Rochester Institute of Technology, Rochester, NY 14623, USA}
\affiliation{Laboratory for Multiwavelength Astrophysics, Rochester Institute of Technology, Rochester, NY 14623, USA}

\begin{abstract}
    In this paper we introduce a new \textsc{python} package, the \textsc{Pulsar Signal Simulator}, or \psrsigsim, which is designed to simulate a pulsar signal from emission at the pulsar, through the interstellar medium, to observation by a radio telescope, and digitization in a standard data format. We use the \psrsigsim to simulate observations of three millisecond pulsars, PSRs~J1744--1134, B1855+09, and B1953+29, to explore the covariances between frequency-dependent parameters, such as variations in the dispersion measure (\DM), pulse profile evolution with frequency, and pulse scatter broadening. We show that the \psrsigsim can produce realistic simulated data and can accurately recover the parameters injected into the data. We also find that while there are covariances when fitting \DM variations and frequency-dependent parameters, they have little effect on timing precision. Our simulations also show that time-variable scattering delays decrease the accuracy and increase the variability of the recovered \DM and frequency-dependent parameters. Despite this, our simulations also show that the time-variable scattering delays have little impact on the root mean square of the timing residuals. This suggests that the variability seen in recovered \DM, when time-variable scattering delays are present, is due to a covariance between the two parameters, with the \DM modeling out the additional scattering delays.

\end{abstract}


\section{Introduction}
\label{sec:introduction}

Precision timing of millisecond pulsars (MSPs) has allowed us to study some of the most extreme astrophysical phenomena, from the equations of state of neutron stars \citep[e.g.,][]{Antoniadis2013, Stovall2018, Cromartie2019} to some of the most rigorous tests of general relativity \citep[e.g.,][]{Kramer2006, Archibald2018, Zhu2019}.
MSPs have also been used to study the properties and dynamics of the interstellar medium \citep[ISM; e.g.,][]{Levin2016, Jones2017, Lam2019, ShapiroAlbert2020}.
Pulsar timing arrays (PTAs) made up of MSPs are used by the North American Nanohertz Observatory for Gravitational Waves \citep[NANOGrav;][]{McLaughlin2013}, the European Pulsar Timing Array \citep[EPTA;][]{Kramer2013}, and the Parkes Pulsar Timing Array \citep[PPTA;][]{Hobbs2013} to search for gravitational waves (GWs) from supermassive black hole binary systems \citep[e.g.,][]{Shannon2013, Zhu2014, Lentati2015, Shannon2015, Arzoumanian2016, Babak2016, Verbiest2016, Arzoumanian2018_GW, Aggarwal2019, Witt2020, NG12yrGW2020}.

For experiments focused on GW detection and characterization, the characterization of noise in the detector is critical \citep{Cordes2010, Cordes2013, Lam2018d, Hazboun2019}.
There are many sources which may contribute to the uncertainty of a pulse time of arrival (TOA), making these detections challenging \citep[e.g.,][]{Lam2018d}.
In particular, various frequency-dependent effects due to both the ISM and the emission at the MSP may increase the uncertainty of a pulse TOA.
These include variations in the dispersion measure (\DM), i.e., the integrated column density of free electrons along the line of sight.
Time delays due to dispersion are $\propto$ \DM $\times$ $\nu^{-2}$, where $\nu$ is the frequency of the radio emission; these variations may result in excess noise if they are not modeled appropriately \citep[e.g.,][]{Jones2017, Lam2018b}. 
Similarly, pulse scatter broadening due to inhomogeneities in the ISM will also cause time-variable delays. Scattering delays are expected to be $\propto \nu^{-4}$ \citep[e.g.,][]{Shannon2012, Lam2019} and will also result in excess noise if not modeled or mitigated.
Finally, evolution of the pulse shape with frequency may also increase the uncertainty of the pulse TOAs if it is not well modeled \citep[e.g.,][]{Kramer1998, Pennucci2014}.

In pulsar timing the time variations in pulsar \DM are often modeled by fitting for a $\Delta$\DM, as an epoch-dependent offset from a fiducial \DM value. The model for \DM variations used in NANOGrav data sets is a piecewise-constant set of offsets, referred to as `\DMX', with a value for each observing epoch \citep[e.g.,][]{Arzoumanian2016, Jones2017, Arzoumanian2018}.
However, accounting for effects such as scattering and profile evolution are more difficult.
To account for profile evolution, Frequency-Dependent, or ``FD", parameters, polynomial coefficients in log-frequency space, along with a JUMP parameter which accounts for additional unmodeled profile evolution and other effects between low and high frequency data, are typically added to the pulsar timing model \citep{Zhu2015, Arzoumanian2016}.

The number of \FD parameters fit varies for each MSP \citep{Arzoumanian2016} but all terms are expected to be covariant with any other frequency-dependent timing parameters, including \DMX and the JUMP parameter.
While it is generally assumed that the largest component of the frequency-dependent time delay accounted for by \FD parameters is due to intrinsic pulse profile evolution with frequency \citep{Zhu2015}, \FD parameters will also account for the average scattering broadening over the course of a data set.

Here we present an analysis of the covariance between the \DMX and \FD parameters, as well as of the contributions of non-$\nu^{-2}$ effects to both the \FD and \DMX parameters using simulated data generated with the \textsc{Pulsar Signal Simulator}\footnote{\url{https://github.com/PsrSigSim/PsrSigSim}} (\psrsigsim) \textsc{python} package \citep{Hazboun2020}.
The \psrsigsim allows us to directly simulate variations in \DM, frequency-dependent pulse profile evolution, and pulse scatter broadening to directly quantify how each of these contributions affects the recovered timing model parameters.
Using simulated data allow us to constrain the impacts of any simulated effects on timing model parameters, precision pulsar timing, and the covariances between the frequency-dependent effects.

We briefly describe the \psrsigsim package in \S \ref{sec:psrsigsim}. In \S \ref{sec:methods} we describe our data analysis pipeline.
Our various simulated data sets are described in \S \ref{sec:sim_data} and the results and analysis of the simulated data are presented in \S \ref{sec:results}.
The implications of our results on precision pulsar timing are presented in \S \ref{sec:implications}.
Finally, we  present concluding remarks and future work in \S \ref{sec:conclusions}.

\section{PsrSigSim Description}
\label{sec:psrsigsim}

The \psrsigsim is a \textsc{python}-based package designed to simulate a realistic pulsar signal including emission at the pulsar, transmission through the ISM, observation by a radio telescope, and output of a data file \citep{Hazboun2020}. Simulations are run on an observation by observation basis and can be run multiple times to create multiple epochs of data.
The \psrsigsim  has a variety of uses for educational purposes \citep{Gersbach:2019}, but here we focus on its use as a scientific simulation tool in this paper.

The package includes modules for various signal classes which define attributes of the signal and observation such as the center frequency, bandwidth, number of frequency channels, and, for the \textsc{filterbanksignal} class which is used in this work, the number of subintegrations and their length.
All \textsc{signal} classes also have an option for the number of polarizations, however the \psrsigsim current only supports total intensity signals, assumed to be the sum of two polarizations.
The \psrsigsim also enables single-pulse simulations using the \textsc{filterbanksignal} class, though not used for this work.

The \textsc{pulsar} class is used to define the properties intrinsic to the pulsar, such as the period ($P$), the mean flux ($S_{\rm{mean}}$) and its reference frequency, and the spectral index ($\alpha$).
In order to define a pulse profile, the \textsc{pulsar} class makes use of either a \textsc{profile} class, for a single profile to be used at all frequency channels, or a \textsc{portrait} class, for a 2-D, frequency-dependent pulse-profile array.
The profiles can be defined in these classes either through the amplitude, position, and width of any number of Gaussians, by defining a function that describes the profile shape as a function of phase, or by supplying a data array representative of the pulse shape.
To define the pulse profile, the \textsc{pulsar} class takes one of these \textsc{profile} or \textsc{portrait} classes.

The \textsc{ism} class is used for modeling the effects of the ISM on the pulsar signal and also account for intrinsic profile evolution.
It includes attributes such as \DM, \FD parameters, and  scattering timescale.
The \textsc{ism} class enables various signal processing techniques, e.g., the shift theorem, to add radio-frequency dependent delays. The \psrsigsim adds these delays to the pulses at specific points of the simulation dependent on astrophysical and efficiency considerations.
Use of Fourier-based techniques allows the \psrsigsim to account for time delays that have time shifts which are fractional in  phase bins.
In the case of scatter broadening, the input scattering timescale is scaled as a function of frequency based on both a user input reference frequency and scaling law exponent.
The \psrsigsim then shifts the profiles directly in time by the resulting delay, or convolves an exponential scattering tail with the input profiles chosen by a user-set flag within the function.

The \textsc{telescope} class encodes the properties of the desired  telescope necessary to compute the radiometer noise and other observing-site specific effects.
A user is able to supply telescope specifications, like the effective area and system temperature. 
Telescope systems can also be defined with specific \textsc{backend} and \textsc{receiver} classes.
The \textsc{receiver} class is currently primarily responsible for defining a bandpass response and calculating the radiometer noise.
The \textsc{backend} class is currently primarily used to inform on the maximum sampling rate of the telescope backend.
As more features are added to the \psrsigsim, such as baseband signal simulation, more features may be added to the \textsc{backend} class as well, such as simulating a polyphase filterbank.
The \psrsigsim comes equipped with pre-defined Arecibo and Green Bank Telescope systems, but additional systems may be added to these, or a new telescope can easily be defined by the user.

The native output of the \psrsigsim is a simulated pulsar signal in the form of a \textsc{numpy} array \citep{van2011numpy}.
However, for this work output in the \psrfits standard was needed in order for software downstream in the analysis pipeline, such as \psrchive, to accept and process the files \citep{Hotan2004, VanStraten2012}. 
To do this, we utilize the \textsc{pulsar data toolbox}\footnote{\url{https://github.com/Hazboun6/PulsarDataToolbox}} (\textsc{pdat}) \textsc{python} package \citep{hazboun_pdat}.
While \textsc{pdat} is not a part of the the \psrsigsim, we include an \textsc{io} class in the \psrsigsim which contains a number of convenience functions.
These use existing \psrfits files as templates to make new files.
Currently, the size of the data array within the template \psrfits file is changed to match the size of the simulated data array, and subsequent metadata, such as the chosen value of DM, is also edited.

The \psrsigsim is designed to simulate one observing epoch of data at a time; by iterating over sets of input parameters it is possible to produce phase-coherent data sets containing multiple observing epochs.
This phase connection is performed by utilizing the \pint\footnote{\url{https://github.com/nanograv/PINT}} pulsar timing software \citep{Luo2020} and an input pulsar ephemeris to replace the polynomial coefficient (POLYCO) values, which predict the pulsar's phase and period using polynomial expansion over a defined time period.
We also note that no binary parameters or delays are currently included in any delay classes or in the creation of the POLYCOs.
If a user desires to create a new \psrfits file from scratch to contain the simulated data, this can be done with a number of currently existing software packages outside of the \psrsigsim such as \textsc{pdat} \citep{hazboun_pdat}, \textsc{astropy.io.fits}\footnote{\url{https://docs.astropy.org/en/stable/io/fits/}} \citep{astropy:2013, astropy}, or \textsc{fitsio}\footnote{\url{https://github.com/esheldon/fitsio}}.

More detailed descriptions as well as examples can be found on the \textsc{readthedocs}\footnote{\url{https://psrsigsim.readthedocs.io/en/latest/readme.html}} page of the \psrsigsim and in \cite{Hazboun2020}.

\section{Methods}
\label{sec:methods}

Here we will describe the general methods used for making and analyzing our simulated data. 
The details of each set of simulated data appear in \S \ref{sec:sim_data}, while here we cover general processes used to produce the data and simulate each of the effects used.
We discuss first the methods used for simulating the data with the \psrsigsim\footnote{\psrsigsim version 1.0.0 is used throughout this work.} and then the methods used to obtain TOAs and fit the different timing parameters.

\subsection{Generating Simulated Data} \label{subsec:sim_methods}

All of our data are simulated using the \psrsigsim \textsc{python} package described in \S \ref{sec:psrsigsim}.
For this work, we look at three different MSPs that are part of the NANOGrav pulsar timing array experiment \citep{Arzoumanian2018}, PSRs~J1744--1134, B1855+09, and B1953+29.
These MSPs span a range of DMs, potentially allowing us to look at the covariances between \DMX and \FD parameters as a function of mean \DM and/or number of \FD parameters.
Additionally they all have notable profile evolution \citep{NANOGravWB}, rather long timing baselines \citep{Arzoumanian2018}, and significant \DM variations \citep{Jones2017}.
For each simulation, we have a set of defined pulsar and observation parameters listed in Tables \ref{tab:pulsar_params} and \ref{tab:backend_params2}.
These include the pulsar's name, period, \DM, mean flux, spectral index, the desired bandwidth of the observation, number of frequency channels, center observing frequency, the observation length, and the telescope name.
To simulate \DM variations, we determine the individual \DMX injected at each epoch using the trends from \cite{Jones2017}, shown in Table \ref{tab:dm_params}.
We use the \DM reported in Table \ref{tab:pulsar_params} as a reference \DM where the injected \DMX is zero.
This reference \DM is taken to be the value at the center epoch of the simulations, and when sinusoidal trends are added it is the value at phase zero.
No additional noise is added to the predictions by these trends.
If any other parameters are desired, such as \FD parameters or scattering timescale ($\tau_{\rm{d}}$), these may also be defined and used in the simulation.

\begin{figure}[t]
    \centering
    \includegraphics[width=9cm]{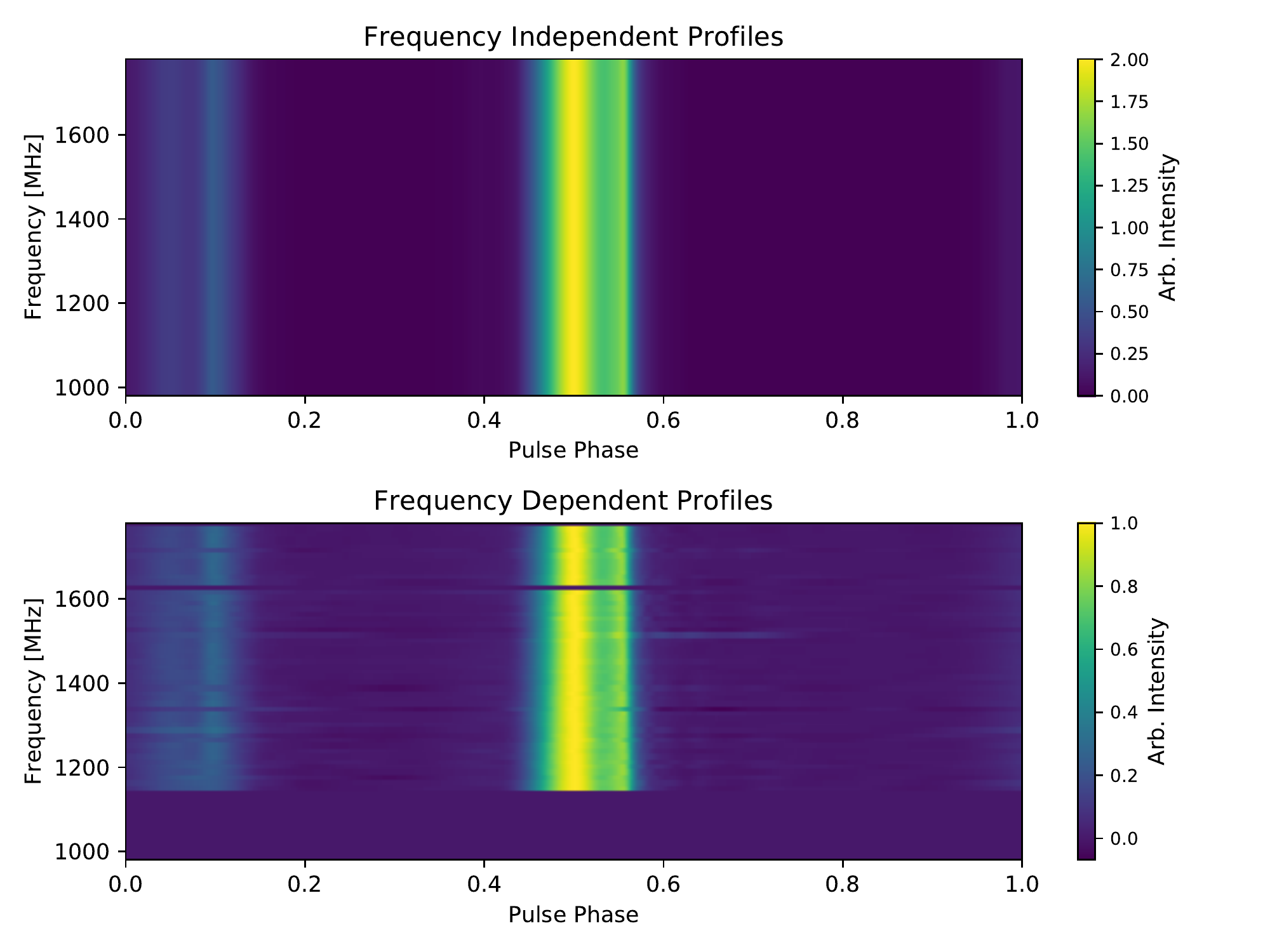}
    \caption{Injected pulse profiles for PSR~B1855+09 used for this work. No additional scattering has been injected, and all peaks are normalized to an arbitrary intensity of one. \textit{Top}: The same pulse profile, here the NANOGrav 11-yr profile for PSR~B1955+09, is used for every frequency channel if no frequency-dependent pulse profile evolution is desired. \textit{Bottom}: Frequency-dependent pulse profiles obtained from modeling a single real observation of PSR~B1855+09. No apparent scintillation or spectral index effects remain after following the process detailed in \S \ref{subsec:sim_methods}, and the frequency-dependent variations are clearly shown. Some channels have been removed due to RFI contamination.}
    \label{fig:Prof_Example}
\end{figure}

We define a pulse shape to be input into the \psrsigsim for each observation depending on the simulated backend and receiver combination to mimic the standard timing procedure described in \cite{Demorest2013} and \cite{Arzoumanian2016}.
For this work, each set of pulse profiles is defined as a 2-D array in frequency and pulse phase, where we use 2048 phase bins.
While the real NANOGrav observations record a different number of frequency channels depending on the receiver-backend combination, either 64, 128, or 512, all are eventually folded down to 64 frequency channels \citep{Arzoumanian2018}.
As we can simulate any number of initial frequency channels, we simulated all of our initial observations with 64 frequency channels to avoid needless post-processing.
For similar reasons, we also simulate all of our data with just a single subintegration of length equal to the total observing length.
If no profile evolution with frequency is desired, we use the NANOGrav 11-yr profile template defined at the appropriate center frequency for the pulse profile.
This is input in the \psrsigsim as a 1-D array \textsc{dataprofile} object, which is then tiled within the \psrsigsim so that the profile is the same in every frequency channel.
An example is shown in the top panel of Figure \ref{fig:Prof_Example}.

When a 2-D array of frequency-dependent profiles is desired as the input into the \psrsigsim, we create them by starting with a post-processed, high signal-to-noise (S/N) NANOGrav observation with 64 frequency channel-depended profiles.
This is done to be sure that all RFI has been removed and the data have been properly calibrated, though each observation was inspected by-eye to confirm this.
We then smoothed this data using the \textsc{psrsmooth} function of the \psrchive data processing package \citep{Hotan2004, VanStraten2012}.
These smoothed profiles are then formatted into a 2-D \textsc{python} data array in frequency and pulse phase, as described above, using the \pypulse\footnote{\url{https://github.com/mtlam/PyPulse}} \textsc{python} package \citep{LamPyPulse}.
Only one set of model profiles was used for each receiver-backend combination. 
For example, if we simulate multiple observations at 1400~MHz, the noise-free profiles used at every simulated observing epoch will be the same, though the pulse shape may change with the observing frequency.
However, the white noise that is added to the simulations will vary from epoch to epoch.

Since we have chosen to model our frequency-dependent profiles using real pulsar data for this work, we must also account for the effects of a pulsar's spectral index \citep[e.g.,][]{Jankowski2018} and scintillation due to the ISM.
Both of these effects are present in all pulsar observations, and if uncorrected, will change the pulse flux as a function of observing frequency in a non-user defined way.
To  remove these intrinsic effects, the pulse profiles were normalized such that all profiles have a peak flux of one in arbitrary flux units.
However, to make our simulated data as realistic as possible a user-defined spectral index, reported in Table \ref{tab:pulsar_params}, is added back into the simulated data when the pulses are created.
To do this, each normalized profile is multiplied by a frequency dependent constant such that
\begin{equation}
     S_{\rm{mean}}(\nu) = S_{\rm{mean}}(\nu_{\rm{ref}}) \left( \frac{\nu}{\nu_{\rm{ref}}} \right)^{\alpha}. 
\end{equation}
Here $S_{\rm{mean}}(\nu_{\rm{ref}})$ is the user-input mean flux referenced to some frequency, $\nu_{\rm{ref}}$, $\nu$ is the center frequency of each frequency channel for each profile, $\alpha$ is the user input spectral index, and $S_{\rm{mean}}(\nu)$ is the new mean flux of the spectral index adjusted profile at a frequency $\nu$.

Since our frequency-depended profiles were created from real, post-processed observations, profiles at some frequency channels had been removed due to contamination by radio frequency interference (RFI).
Since we cannot realistically model profiles in the frequency channels that have been removed, we instead replace them with a profile of zeros.
When creating TOAs from these profiles, all channels that were replaced with zeros in this way were removed as well, and are not included in any pulsar timing model fitting (described in \S \ref{subsec:toas_and_res}.
This 2-D array of frequency-dependent profiles is then input into the \psrsigsim as a \textsc{dataportrait} object, and an example is shown in the bottom panel of Figure \ref{fig:Prof_Example}.
We note these were all choices made for this work, and that the \psrsigsim is capable of using any set of 1- or 2-D user-generated pulse profiles.

While there is some inherent scatter broadening already contained within the real data used for our model frequency-dependent profiles, we do not know a priori how much the profiles have been scatter broadened, and hence can not separate this effect from intrinsic profile evolution.
In some of our simulations (described in \S \ref{sec:sim_data}), however, we simulate pulse scatter broadening using the \textsc{ism} class by defining a single input $\tau_{\rm{d}}$, referenced to an initial input frequency,  for each simulated epoch.
Within the \psrsigsim, $\tau_{\rm{d}}$ is scaled for each frequency channel as
\begin{equation} \label{eq:tau_d_scale}
    \tau_{\rm{d}_{i}} = \tau_{\rm{d}} 
    \left( \frac{\nu_{i}}{\nu_{\rm{ref}}} \right)^{\beta}.
\end{equation}
Here $\nu_{\rm{ref}}$ is the reference frequency of the input $\tau_{\rm{d}}$, $\nu_{i}$ is the center frequency of the $i$th frequency channel, and $\beta$ is the scaling law exponent.
The exponential scattering tail for each frequency channel is then calculated as $\exp(-t/\tau_{\rm{d}_{i}})$, where $t$ is the fractional time of each profile bin.
The resulting frequency-dependent exponential scattering tails are then convolved with the pulse profiles.
For our simulations, we assume a Kolmogorov medium, so $\beta=-4.4$, though $\beta$ can also be set by the user within the \psrsigsim.
While it has been found that measurements of $\beta$ deviate from a Kolmogorov medium \citep[e.g.][]{Levin2016, Turner2020}, we have chosen to use a constant value to minimize the number of variables that affect the covariance between \DMX, \FD, and $\tau_{\rm{d}}$.
While studying how varying $\beta$ may affect these covariances is certainly of interest, this added complexity is beyond the scope of this work.

For this work, we have chosen to run simulations with both a single value of $\tau_{\rm{d}}$ across all epochs and with time-varying $\tau_{\rm{d}}$.
In the case of a time-varying $\tau_{\rm{d}}$, we have chosen input values of $\tau_{\rm{d}}$ by randomly sampling a Gaussian distribution with mean and 1$\sigma$ variation reported in Table \ref{tab:pulsar_params} and then taking the absolute value of the sampled $\tau_{\rm{d}}$.
However, for PSR~B1953+29, no RMS variation for $\tau_{\rm{d}}$ was reported by \citet{Levin2016}, so we use a 1$\sigma$ variation of 20\% the mean value.
A different value of $\tau_{\rm{d}}$ is then input for every simulated epoch of observations.
We note that because the \psrsigsim simulates just a single epoch at a time, a user may choose input values for $\tau_{\rm{d}}$ using any method.

After the pulses are simulated, they are dispersed with the \textsc{ism} class.
This is done by calculating the time delay due to dispersion,
\begin{equation}
 \label{eq:DM_delay}
    \Delta t_{\DM} = 2.41\times10^{-4}~\mathrm{s} \left(\frac{\DM}{\rm{pc}~\rm{cm}^{-3}}\right)  \left(\frac{\nu}{\rm{MHz}}\right)^{-2},
\end{equation}
in each frequency channel with respect to infinite frequency.
Here $\nu$ is the center frequency of each frequency channel.
The pulses are then shifted in Fourier space \citep{Bracewell1999} to account for time shifts that are fractional sizes of the discrete time bins.
For this work, the \DM used is the sum of the base value reported in Table \ref{tab:pulsar_params} and the individual \DMX determined at each epoch as described above.
However, we note that in general, the user may input any desired \DM into the \psrsigsim.

Non-dispersive frequency-dependent time delays are also simulated. 
In particular, we directly shift the pulses in time to simulate the ``\FD'' model for frequency-dependent pulse profiles.
To do this, we calculate $\Delta t_{\FD}$ as \citep{Zhu2015, Arzoumanian2016}
\begin{equation} \label{eq:FD_delay}
    \Delta t_{\FD} = \sum_{i=1}^{n} c_{i} \ln\left({\frac{\nu}{1~\rm{GHz}}}\right)^{i}.
\end{equation}
Here $c_{i}$ are the polynomial coefficients in time units, more often referred to as the \FD parameters, such that $c_{1}=\FD1$ and so on, $n$ is the number of coefficients, and $\nu$ is the center frequency of each frequency channel.
The pulses are then shifted in Fourier space as with $\Delta t_{\DM}$.
Within the \psrsigsim the \FD parameters are input in units of seconds.
We report the number and value of each \FD parameter used for each simulated pulsar in this work in Table \ref{tab:pulsar_params}, though in general the user may input any number of \FD parameters with any value into the \psrsigsim.

Once these delays are added, we then define the \textsc{telescope} used in this work as either the 305-m William E. Gordon Telescope of the Arecibo Observatory or the 100-m Green Bank Telescope of the Green Bank Observatory.
We do this using the default \textsc{arecibo} or \textsc{gbt} definition in the \psrsigsim, though a user may define any telescope system they wish for their own simulations.
Radiometer noise is then added to the simulated data based on the desired receiver-backend configuration.
The noise is sampled from a chi-squared distribution with a number of degrees of freedom equal to the number of single pulses in each subintegration.
This is then multiplied by the noise variance ($\sigma_{S}$), calculated as defined in \cite{Handbook},
\begin{equation} \label{eq:radiometer}
    \sigma_{S} = \frac{\rm{T_{sys}+\rm{T_{sky}}}}{G \sqrt{n_{\rm{p}}~dt~\rm{BW}_{\rm{chan}}}},
\end{equation}
where $\rm{T_{sys}}$ is the system temperature, $\rm{T_{sky}}$ is the sky temperature, $G$ is the telescope gain, $n_{\rm{p}}$ is the number of polarizations, $dt$ is the length of each phase bin or $1/$(sample rate), and $\rm{BW}_{\rm chan}$ is the bandwidth of a frequency channel.
Currently the simulator does not model the sky temperature, so we take $\rm{T_{sky}}=0$ for all simulations.
Since only total intensity signals are supported at this time, we assume that the total intensity is the sum of two intensities and so $n_{\rm{p}}=2$ for all simulations.

Since the user input profiles are normalized within the \psrsigsim, as they may be input with arbitrary units, this is then scaled by the maximum flux, $S_{\rm{max}}$, calculated from the mean flux,
\begin{equation} \label{eq:s_max}
    S_{\rm{max}} = \frac{S_{\rm{mean}} n_{\rm{bins}}}{\sum_{i = 1}^{n_{\rm{bins}}} p_{i}}.
\end{equation}
Here $n_{\rm{bins}}$ are the number of phase bins per profile (2048 in all of the simulations in this work), and $p_{i}$ is the intensity of the model profile at the $i$th phase bin. 
If using frequency-dependent pulse profiles, the profile with the maximum integrated flux (in arbitrary units) is used.
This radiometer noise is then further scaled by a normalization coefficient, $U_{\rm{scale}}$, since, as mentioned above, the model profiles are normalized within the \textsc{pulsar} class.
This constant is calculated as defined in \cite{Lam2018d},
\begin{equation} \label{eq:U_scale}
    U_{\rm{scale}} = \frac{1.0}{\left(\sum_{i = 1}^{n_{\rm{bins}}} p_{i} \right) / n_{\rm{bins}}},
\end{equation}
where again the profile used is from the frequency channel that results in the maximum integrated flux.

The final simulated data are contained within a \textsc{numpy} array \citep{van2011numpy}. 
However, for this work, since we require the use of the \psrchive software, we have used the convenience functions provided in the \textsc{io} class and described in \S \ref{sec:psrsigsim} to save the full simulated data array as a \psrfits file \citep{Hotan2004} as described in \S \ref{sec:psrsigsim}.

\subsection{TOAs and Residuals} \label{subsec:toas_and_res}

Once the data have been simulated in \psrfits file format, they are analyzed with both \psrchive and \pint \footnote{\pint version 0.7.0 is used throughout this work.}.
The data are simulated such that all observations match the post-processed NANOGrav standard timing methods \citep{nanopipe}, with a single subintegration and 64 frequency channels.
For simulated Arecibo data, this results in frequency channels with widths of 1.5625 and 12.5~MHz at 430 and 1400~MHz, respectively.
For simulated GBT data, this results in frequency channels with widths of 3.125 and 12.5~MHz at 820 and 1400~MHz, respectively.

TOAs are obtained from the simulated data with the \textsc{pat} function in \psrchive.
We use the corresponding NANOGrav 11-yr pulse profile templates for the template matching process.
This method employs a constant template profile at the appropriate frequency bands regardless of whether frequency-dependent profiles were used in the simulations to better match the standard template-fitting methods used by NANOGrav \citep{Taylor1992, nanopipe}.

Normally, certain frequency channels are ignored in the NANOGrav timing pipeline as they are highly contaminated by RFI \citep{Arzoumanian2016, nanopipe}. 
While we generate no RFI in our simulated data, we mimic this loss in sensitivity by removing all TOAs from these ranges in all simulation analyses.
This includes channels where no frequency-dependent profile model has been generated, as described above and shown in Figure \ref{fig:Prof_Example}.
For simulated Arecibo data, the removed ranges are 380--423, 442--480, 980--1150, and 1618--1630~MHz.
For simulated GBT data, the removed ranges are 794.6--798.6, 814.1--820.7, 1100--1150, 1250--1262, 1288--1300, 1370--1385, 1442--1447, 1525--1558, 1575--1577, 1615--1630~MHz.

We then calculate the timing residuals using the \pint pulsar timing package \citep{Luo2020}.
Each pulsar timing model is extremely simple and includes only the position, period, \DM, \DMX, the number of \FD parameters equal to that listed in \cite{Arzoumanian2018}, and one JUMP parameter to account for unmodeled profile evolution and other effects between the low and high frequency simulated data.
Of these we fit only combinations of \DMX, \FD, and JUMP parameters, holding all other values fixed.
Since we have not included any motions of the Earth, we assume that all TOAs that we have obtained are already barycentered.
We do not include any effects such as parallax, proper motion, or binary motion and therefore do not fit for these in our timing model.

When fitting the different \DMX values for each simulation, we follow \cite{Arzoumanian2016} and \cite{Arzoumanian2018} and bin our simulated TOAs in groups of 15~days for simulated epochs before MJD~56000, and 6~days after MJD~56000.
The adjustment in binning comes from the less frequent observations that occurred early on in the NANOGrav timing program \citep{Arzoumanian2018}.
We then fit our timing model parameters using the generalized least squares fitter in \pint and compare the fit values of \DMX and the \FD parameters, which we will denote as $\widehat{\DMX}$ and $\widehat{\FD}$, to the injected values.

\section{Simulated Data}
\label{sec:sim_data}

The parameters that were used to make the simulated data for PSRs~J1744--1134, B1855+09, and B1953+29 are shown in Table \ref{tab:pulsar_params}.
We note that while real pulsars have many additional timing parameters (e.g., spin down, proper motions, etc.), we do not simulate these effects in any of our data sets presented here.
Our simulated data therefore represent barycentered observations that have had all non-frequency-dependent delays removed\footnote{The inclusion of additional timing parameters and the covariances between them is generally of interest and is a topic for future work.}.

\begin{deluxetable}{cccc}
\tablecaption{Simulated Pulsar Parameters \label{tab:pulsar_params}}
\tablecolumns{4}
\tablehead{
\colhead{Parameter} & \colhead{J1744$-$1134} & \colhead{B1855+09} & \colhead{B1953+29}
}
\startdata
Period (ms) & 4.075 & 5.362 & 6.133 \\
DM (pc cm$^{-3}$) & 3.09 & 13.30 & 104.5 \\
FD1 ($\upmu$s) & $-383.4 \pm 88.5$ & $128.7 \pm 24$ & $139.5 \pm 7.8$ \\
FD2 ($\upmu$s) & $395.6 \pm 89.5$ & $-147.5 \pm 29$ & $-61.0 \pm 7.8$ \\
FD3 ($\upmu$s) & $-241.1 \pm 60.2$ & $81.6 \pm 18$ & - \\
FD4 ($\upmu$s) & $98.9 \pm 23.3$ & - & - \\
$S_{430}$ (mJy) & - & 14.56 & 10.77 \\
$S_{820}$ (mJy) & 2.93 & - & - \\
$S_{1400}$ (mJy) & 0.98 & 2.13 & 0.69 \\
$\alpha$ & $-1.77$ & $-1.45$ & $-2.16$ \\
$\tau_{\rm{d}}$ (ns) & $3.3 \pm 1.6$ & $8.1 \pm 4.4$ & $55.3 (\pm 11.1)$ \\
\enddata
\tablecomments{Parameters describing the three MSPs that were used in these simulations. Period, DM, and FD1-4 values are from the NANOGrav 11-yr data set \citep{Arzoumanian2018}. Flux values ($S$) and spectral index ($\alpha$) values are from \cite{Arzoumanian2020}.} Scattering timescales $\tau_{\rm{d}}$ are all referenced to 1500~MHz and for PSRs~J1744$-$1134 and B1855+09 come from \cite{Turner2020}, and for PSR~B1953+29 from \cite{Levin2016}. All uncertainties are $1\sigma$, with the uncertainties on scattering delay defined as the RMS variation over the data set. No variation on $\tau_{\rm{d}}$ for PSR~B1953+29 was reported in \cite{Levin2016} so we define it to be 20\% the measured value.
\end{deluxetable}

The simulated data sets for each pulsar are split into two sets of simulations.
The first set consists of simulations where the pulse profile is frequency-independent for a given observing band.
The recovered parameters generally match  the injected parameters in this set and are used primarily for comparison.
The second set uses realistic, frequency-dependent pulse profiles as described in \S \ref{subsec:sim_methods}.
In total, we simulated nine different data sets with different injections for each pulsar.
Five of them used frequency-independent pulse profiles for comparison purposes, and the other four were used to analyze the covariances between the frequency-dependent parameters.
The basic injections and values used for each simulation can be found in Table \ref{tab:sims_description}.

All simulations span the same length as the observations of each pulsar using the NANOGrav observing epochs from \cite{Arzoumanian2018}.
While it is not necessary to simulate a full data set like this to explore the covariances between these parameters, we do this in part to demonstrate that the \psrsigsim can simulate long sets of unevenly samples observing epochs while maintaining a precise phase connection.
In addition, it demonstrates the efficiency of the \psrsigsim, as the total time for each of the nine simulations to run for the two longer data sets, those of PSRs~J1744--1134 and B1855+09, was $\sim5$~minutes on an Intel(R) Xeon(R) CPU E5-2630 0 @ 2.30GHz with 24 processors and 64 GB of RAM.
In addition to using realistic observing epochs as a benchmarking test for the \psrsigsim, having many DMX bins better quantifies the variations that may be seen in DMX due to profile evolution or scatter broadening.
Finally, as the FD parameters are fit globally over the entire data set, their fit values are sensitive to the length and number of observations in the simulated data set.

For simulated observations using the Arecibo telescope, we simulate only data from the PUPPI backend \citep{Ford2010} and for observations simulated using the GBT, we simulate only data using the GUPPI backend \citep{DuPlain2008}.
While much of the early observations of these MSPs were done with the GASP or ASP backend \citep{Demorest2007}, there are additional systematics introduced into the pulsar timing when switching between backends that is beyond the scope of this work.
The observing frequencies of each pulsar, either 430~MHz or 1400~MHz at Arecibo or 820~MHz or 1400~MHz at the GBT, are the same as in \cite{Arzoumanian2018}.
The parameters for each receiver-backend
combination are reported in Table \ref{tab:backend_params2}.
The observation lengths, the time of each simulated observing epoch, come from the length of the actual observation that was used to generate the frequency-dependent pulse profiles.
These lengths represent a typical observing length for each pulsar as observed by NANOGrav, though they are kept constant for each simulated observing epoch. 


\begin{deluxetable*}{cccccccc}
\tablecaption{Description of Backend/Receiver Parameters \label{tab:backend_params2}}
\tablecolumns{9}
\tablehead{
\colhead{PSR} & \colhead{Backend} & \colhead{Receiver} & \colhead{MJD Range} & \colhead{Center Frequency} & \colhead{Bandwidth} & \colhead{Frequency Channels} & \colhead{Observation Length} \\
\colhead{} & \colhead{} & \colhead{} & \colhead{} & \colhead{(MHz)} & \colhead{(MHz)} & \colhead{} & \colhead{(s)} 
}
\startdata
J1744$-$1134 & GUPPI & 820 & 53217-57369 & 820 & 200 & 64 & 1578 \\
J1744$-$1134 & GUPPI & L-wide & 53216-57367 & 1500 & 800 & 64 & 1742 \\
B1855+09 & PUPPI & 430 & 53358-57375 & 430 & 100 & 64 & 1204 \\
B1855+09 & PUPPI & L-wide & 53358-57375 & 1380 & 800 & 64 & 1270 \\
B1953+29 & PUPPI & 430 & 55760-57348 & 430 & 100 & 64 & 1443 \\
B1953+29 & PUPPI & L-wide & 55760-57376 & 1380 & 800 & 64 & 1440 \\
\enddata
\tablecomments{Parameters used for each simulated backend. MJD ranges, center frequencies, bandwidths, taken from the NANOGrav 11-yr data set \citep{Arzoumanian2018}. All simulated observations were simulated with 64 frequency channels. Observation lengths are from the observations which were used to obtain high S/N pulse profile templates.}
\end{deluxetable*}

\begin{deluxetable}{cccc}
\tablecaption{DM Variation Parameters \label{tab:dm_params}}
\tablecolumns{4}
\tablehead{
\colhead{PSR} & \colhead{DM Slope} & \colhead{DM Amplitude} & \colhead{DM Period}\\
\colhead{} & \colhead{($10^{-3}$ pc cm$^{-3}$ yr$^{-1}$)} & 
\colhead{($10^{-4}$ pc cm$^{-3}$)} & \colhead{(days)} 
}
\startdata
J1744$-$1134 & $-0.069$ & 0.4 & 383 \\
B1855+09 & 0.382 & 0.5 & 364  \\
B1953+29 & $-1.3$ & 3.0 & 356 \\
\enddata
\tablecomments{Slope, amplitude, and period of the DM variations used for each pulsar that was simulated as derived by \cite{Jones2017}.}
\end{deluxetable}

\begin{center}
\begin{deluxetable*}{Lccccc}
\tablecaption{Description of Simulated Data sets \label{tab:sims_description}}
\tablecolumns{6}
\tablehead{
\colhead{Simulation Name} & \colhead{DM Variations} & \colhead{FD Injection} & \colhead{Profile Evolution} & \colhead{Constant Scatter Broadening} & \colhead{Time-Variable Scatter Broadening}
}
\startdata
No Variations & N & N & N & N & N \\
\hline
DM Variations & Y & N & N & N & N \\
\hline
FD Injections & N & Y & N & N & N \\
\hline
Time-Variable Scatter Broadening w/ Constant DM & N & N & N & N & Y \\
\hline
DM \& FD Injections & Y & Y & N & N & N \\
\hline
Profile Evolution & N & N & Y & N & N \\
\hline
DM \& Profile Evolution & Y & N & Y & N & N \\
\hline
Scatter Broadening & Y & N & Y & Y & N \\
\hline
Time-Variable Scatter Broadening & Y & N & Y & N & Y \\
\enddata
\tablecomments{Description of parameters included in each simulated data set. A `Y' means the parameter was injected into the simulation, while `N' means it was not injected. If DM variations are included, the slope, amplitude, and period of the DM variations are presented in Table \ref{tab:dm_params}. If FD injection is included the profiles used do not vary in frequency and are instead direction shifted in time by the delay described the NANOGrav 11-yr FD parameters. If profile evolution is used, then frequency-dependent profile models based on actual observations of each pulsar are used for the injected pulse profiles. If scatter broadening is used, then the injected profiles are convolved with an exponential described by a scattering timescale, $\tau_{\rm{d}}$. If constant, then each epoch uses the mean value of $\tau_{\rm{d}}$ from Table \ref{tab:pulsar_params}, if time-variable, $\tau_{\rm{d}}$ is drawn from a Gaussian distribution with a mean and standard deviation from Table \ref{tab:pulsar_params}.}
\end{deluxetable*}
\end{center}

\subsection{Frequency-Independent Pulse Profile Simulations} \label{subsec:Initial_sims}

\begin{figure*}[]
    \centering
    \includegraphics[width=\textwidth,height=22.5cm,keepaspectratio]{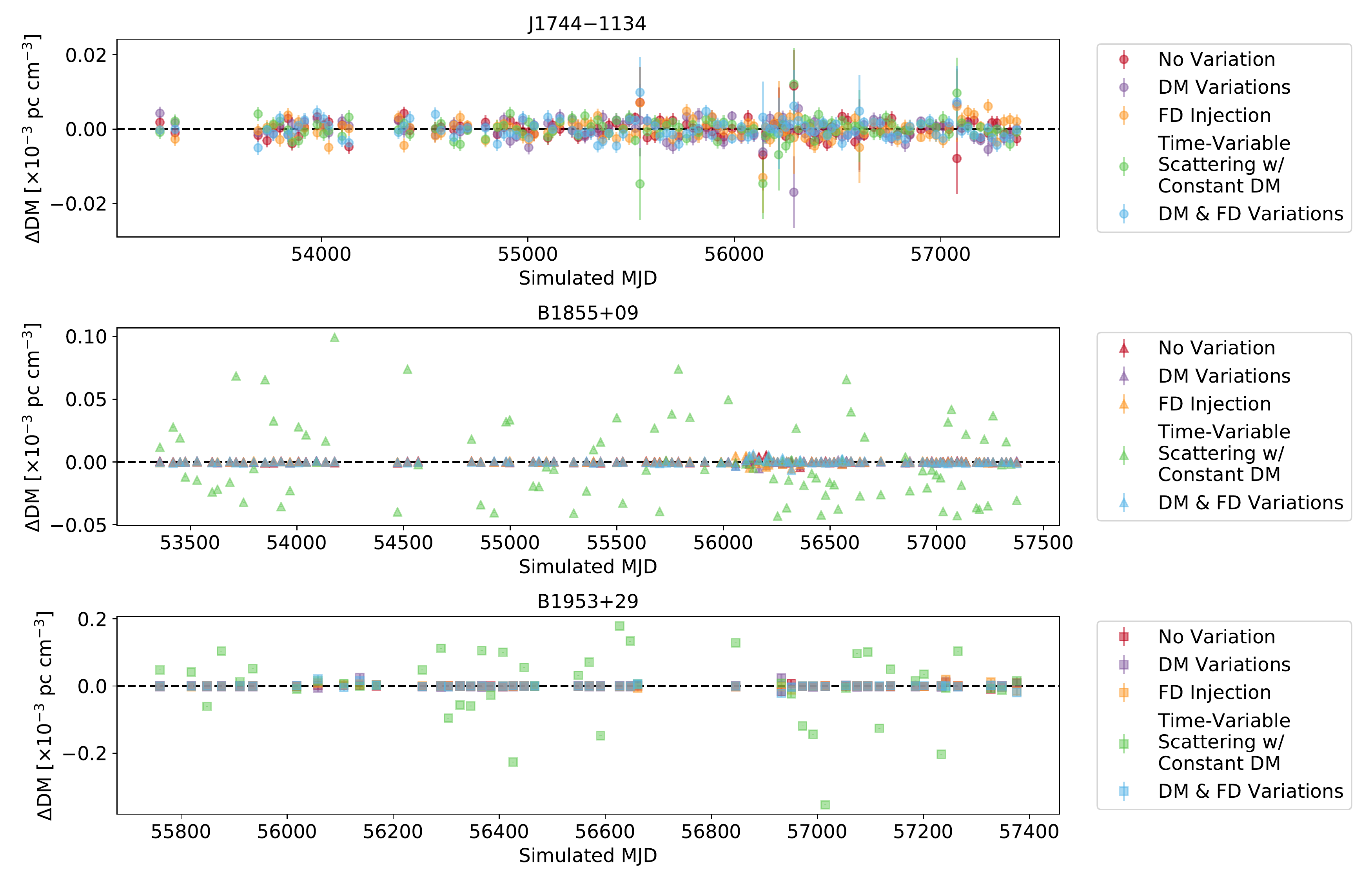}
    \caption{Resulting $\Delta\DM$ values for all three simulated pulsars for simulations where no frequency-dependent pulse profiles were used (described in \S \ref{subsec:Initial_sims}) when fitting for \DMX, all \FD parameters, and a JUMP. Different symbols are used for each pulsar. The black dashed lines represent the zero lines. All points for each pulsar and each simulation are scattered around this zero line, showing that they are being appropriately recovered and fit for.}
    \label{fig:PINT_DMX_Diffs_1}
\end{figure*}

The first simulation in this set was the ``No Variation" simulation.
Every simulated epoch used a constant value of \DM (e.g., all \DMX values were zero), and included no additional time delays (e.g., from \FD parameters or scatter broadening).
The purpose of this was to test the simplest case simulation and provide a baseline to compare to other simulations with addition injected effects.
For each observation we determine $\Delta\DM$, the recovered \DMX minus the injected \DMX, $\Delta\DM=\widehat{\DMX}-\DMX$. 
This is shown in red at every simulated observing epoch for all three simulated pulsars in Figure \ref{fig:PINT_DMX_Diffs_1}.
We note that the $\widehat{\DMX}$ values have been mean subtracted as $(\widehat{\DMX}_{i}-\langle\widehat{\DMX}\rangle)$ for each \DMX epoch $i$, and the error bars shown represent the errors on the mean subtracted value.
This is done because it allows us to separate the uncertainty of each \DMX measurement from that of the mean \DM, since there is a large covariance between these parameters \citep{Arzoumanian2016,Arzoumanian2018}.

\begin{figure}[h]
    \centering
    \includegraphics[width=9cm]{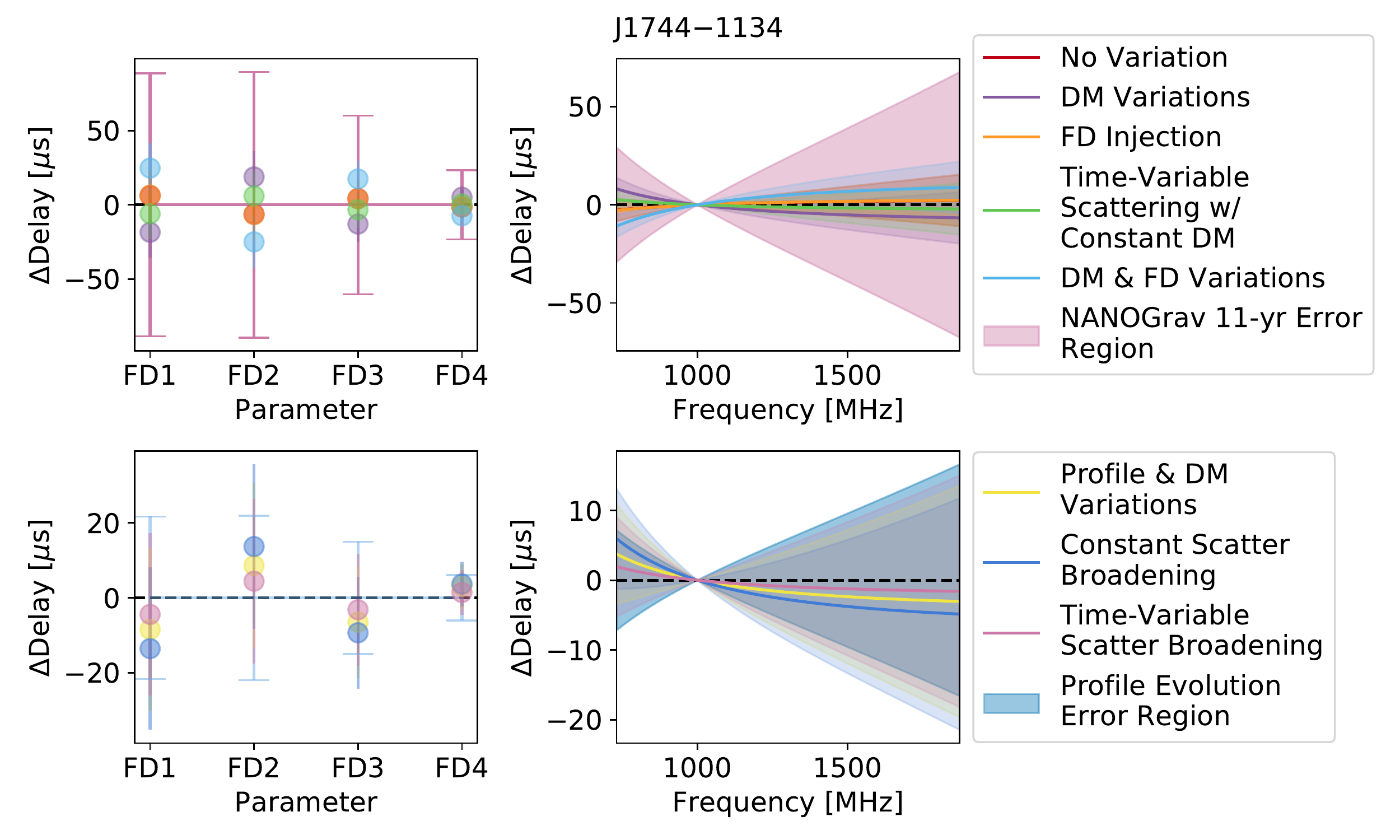}
    \caption{\textit{Upper Left:} Resulting $\Delta\FD$ parameters for five different simulated data sets of eleven years where no frequency-dependent pulse profiles were used for PSR~J1744--1134 (described in \S \ref{subsec:Initial_sims}). All recovered values come from fitting for \DMX, all \FD parameters, and a JUMP. The recovered parameters match well with the injected parameters. \textit{Upper Right:} Same as upper left but showing the delay curve across frequency space defined by the $\Delta\FD$ parameters shown in the upper left plot. Shaded regions represent the 1$\sigma$ recovered errors. \textit{Lower Left:} Same as upper left, but for four different simulations where frequency-dependent pulse profiles were used (described in \S \ref{subsec:Interesting_Sims}). \textit{Lower Right:} Same as upper right, but for the simulations listed in the lower left.}
    \label{fig:PINT_FD_J1744}
\end{figure}

\begin{figure}[h]
    \centering
    \includegraphics[width=9cm]{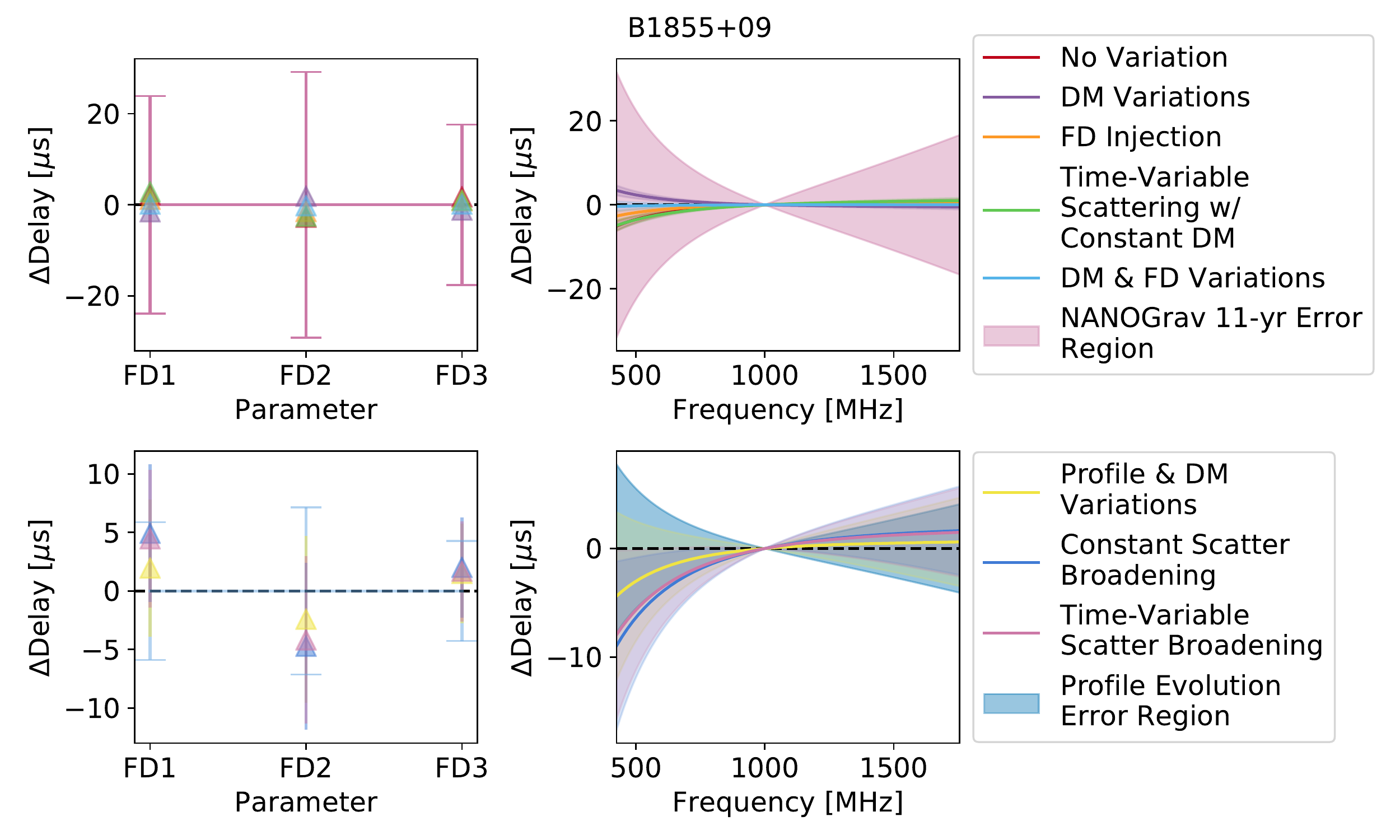}
    \caption{\textit{Upper Left:} Resulting $\Delta\FD$ parameters for five different simulated data sets of eleven years where no frequency-dependent pulse profiles were used for PSR~B1855+09 (described in \S \ref{subsec:Initial_sims}). All recovered values come from fitting for \DMX, all \FD parameters, and a JUMP. The recovered parameters match well with the injected parameters. \textit{Upper Right:} Same as upper left but showing the delay curve across frequency space defined by the $\Delta\FD$ parameters shown in the upper left plot. Shaded regions represent the 1$\sigma$ recovered errors. \textit{Lower Left:} Same as upper left, but for four different simulations where frequency-dependent pulse profiles were used (described in \S \ref{subsec:Interesting_Sims}). \textit{Lower Right:} Same as upper right, but for the simulations listed in the lower left.}
    \label{fig:PINT_FD_B1855}
\end{figure}

\begin{figure}[h]
    \centering
    \includegraphics[width=9cm]{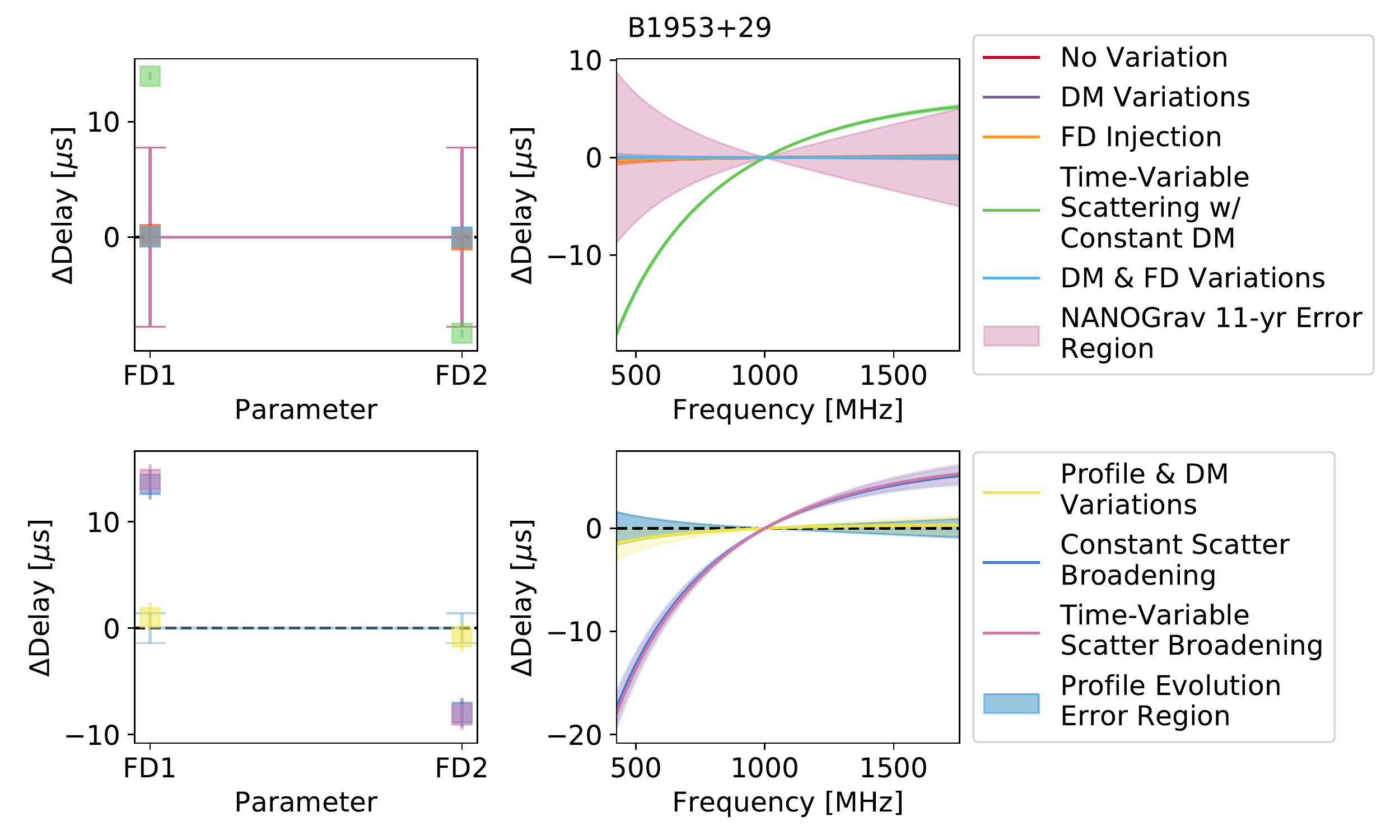}
    \caption{\textit{Upper Left:} Resulting $\Delta\FD$ parameters for five different simulated data sets of eleven years where no frequency-dependent pulse profiles were used for PSR~B1953+29 (described in \S \ref{subsec:Initial_sims}). All recovered values come from fitting for \DMX, all \FD parameters, and a JUMP. The recovered parameters match well with the injected parameters. \textit{Upper Right:} Same as upper left but showing the delay curve across frequency space defined by the $\Delta\FD$ parameters shown in the upper left plot. Shaded regions represent the 1$\sigma$ recovered errors. \textit{Lower Left:} Same as upper left, but for four different simulations where frequency-dependent pulse profiles were used (described in \S \ref{subsec:Interesting_Sims}). \textit{Lower Right:} Same as upper right, but for the simulations listed in the lower left.}
    \label{fig:PINT_FD_B1953}
\end{figure}

We similarly determine $\Delta\FD$ parameters, the recovered \FD parameter minus injected \FD parameter, $\Delta{\FD}_{\rm{i}} = \widehat{\FD_{\rm{i}}}-{\FD}_{i}$, for each individual \FD parameter i.
For this simulation, these shown in red in the two upper panels of Figures \ref{fig:PINT_FD_J1744}, \ref{fig:PINT_FD_B1855}, and \ref{fig:PINT_FD_B1953} for each pulsar respectively.
All of the recovered $\Delta\DM$ and $\Delta\FD$ parameters for these simulations are shown in the same panels of the same Figures, though with different colors.
All recovered values, $\widehat{\DMX}$ and $\widehat{\FD}$, shown in these figures were determined by fitting for all parameters: \DMX, all \FD parameters, and a single JUMP parameter.

The second simulation is the ``\DM Variations" simulation. 
Here the total injected \DM is the initial value given in Table \ref{tab:pulsar_params} plus a small variation added based on the parameters given in Table \ref{tab:dm_params}.
The variations for all simulated pulsars had both a linear and sinusoidal trend with slope, amplitude, and period as determined by \cite{Jones2017}.
The resulting $\Delta\DM$ and $\Delta\FD$ parameters, similar to that shown for the ``No Variations" simulation, are shown in purple.

The third simulation in this set is the ``\FD Injection."
While the physical process that \FD parameters describe is mainly attributed to pulse profile evolution in frequency \citep{Zhu2015}, they define a time delay directly given by Eq. \ref{eq:FD_delay}.
To provide a baseline for recovering the injected \FD parameters, we directly shift the simulated pulses in time based on the \FD parameters listed in Table \ref{tab:pulsar_params}.
We do this instead of varying the profiles directly because we do not know a priori what the shifts due to profile evolution are, we can only fit them empirically.
The resulting $\Delta\DM$ and $\Delta\FD$ parameters for this simulation are shown in orange.

The fourth simulation here is ``Time-Variable Scattering with Constant \DM."
Here we again use a constant value of \DM, but also inject time-variable values of $\tau_{\rm{d}}$ on a per-epoch basis, selected as described in \S \ref{subsec:sim_methods}.
This gives us baseline to compare how $\widehat{\DMX}$ is affected by this time-variable scattering in more complex simulations.
The resulting $\Delta\DM$ and $\Delta\FD$ parameters for this simulation are shown in light green.

Our final initial simulation, ``\DM \& \FD Injections" is a combination of the second and third initial simulations.
This was done to provide a baseline for the accuracy of $\widehat{\DMX}$ and $\widehat{\FD}$ since they are both dependent on the emission frequency.
The resulting $\Delta\DM$ and $\Delta\FD$ parameters for this simulation are shown in light blue.

While all of the recovered values shown in Figures \ref{fig:PINT_DMX_Diffs_1}, \ref{fig:PINT_FD_J1744}, \ref{fig:PINT_FD_B1855}, and \ref{fig:PINT_FD_B1953} come from fitting for all parameters, \DMX, all \FD parameters, and a JUMP, we also fit each of these simulations using just a single JUMP, just \DMX and a single JUMP, and just all applicable \FD parameters and a single JUMP (for a total of four different fits for each simulation).
We report the RMS of the timing residuals ($\cal{R}_{\rm{rms}}$), reduced chi-squared ($\chi^{2}_{\rm{r}}$ of the fit timing model, the RMS of the $\Delta\DM$ values ($\widehat{\DMX}-\DMX$), $\Delta\DM_{\rm{rms}}$, the RMS of the $\Delta\FD$ parameters ($\widehat{\FD_{\rm{i}}}-\FD_{i}$), $\Delta\FD_{\rm{rms}}$, and the fit value of the JUMP, for each of these fits per simulation per pulsar in Tables \ref{tab:J1744-1134_fit_results}, \ref{tab:B1855+09_fit_results}, and \ref{tab:B1953+29_fit_results} respectively.

\begin{figure*}[]
    \centering
    \includegraphics[width=\textwidth,height=22.5cm,keepaspectratio]{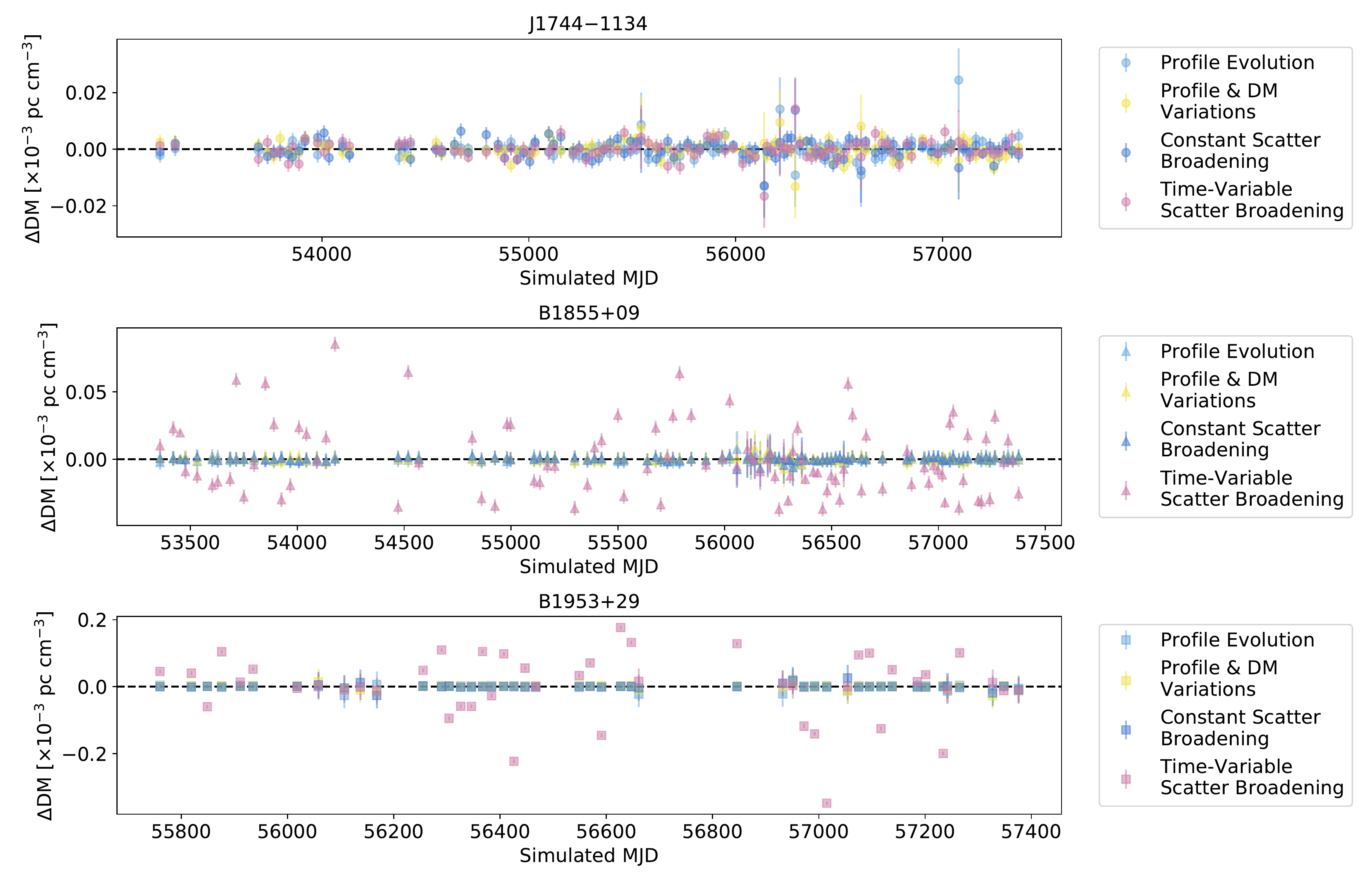}
    \caption{Resulting $\Delta\DM$ values for all three simulated pulsars for simulations where frequency-dependent profile evolution was modeled (described in \S \ref{subsec:Interesting_Sims}) when fitting for \DMX, all \FD parameters, and a JUMP. Different symbols are used for each pulsar. The black dashed lines represent the zero lines. All points for each pulsar and each simulation are scattered around this zero line, showing that they are being appropriately recovered and fit for.}
    \label{fig:PINT_DMX_Diffs_2}
\end{figure*}

\subsection{Frequency-Dependent Pulse Profile Simulations} \label{subsec:Interesting_Sims}

Next we use a different set of frequency-dependent profiles, one for each different receiver-backend combination for each pulsar, is used as described in \S \ref{sec:methods}.
As noted in \S \ref{subsec:sim_methods}, we use only one set of frequency-dependent profiles for each receiver-backend combination.
Since there are no variations in the profile evolution in time, e.g., due to scintillation \citep{Cordes1986}, we do not expect to recover exactly the same \FD parameters as reported in Table \ref{tab:pulsar_params}.
We do, however, expect similar \FD parameters, with the same signs and orders of magnitude.

To determine what the contribution of the frequency-dependent profiles is to the \FD parameters, our first simulation in this set, labeled ``Profile Evolution", uses a constant \DM such that all injected \DMX values are zero, and includes only the frequency-dependent profile.
This allows us to determine what the expected contribution of the chosen set of frequency-dependent profiles is, and help to quantify any deviations in $\widehat{\FD}$ as more frequency-dependent effects are added.
The resulting $\Delta\DM$, similar to that shown for the previous set of simulations, are shown in blue in Figure \ref{fig:PINT_DMX_Diffs_2} and the resulting $\Delta\FD$ parameters for this simulation, also in blue, are shown in the two lower panels of Figures \ref{fig:PINT_FD_J1744}, \ref{fig:PINT_FD_B1855}, and \ref{fig:PINT_FD_B1953} for each pulsar respectively.
All values of $\Delta\DM$ and $\Delta\FD$ parameters for these simulations are shown in the same panels of the same Figures, though with different colors.
The resulting $\Delta\FD$ parameters for this simulation have values of zero with an associated error bar, since we do not know their value a priori. For this simulation $\widehat{\FD}$ are used as the baseline for all other simulations in this set.

The second simulation, ``\DM \& Profile Evolution", uses both the frequency-dependent profiles as well as the \DM variations that were used in the initial simulations described in \S \ref{subsec:Initial_sims}. 
This simulation allows us to explore the covariances between \DMX and profile evolution, and compare them to the covariances when \FD parameters are directly injected via time shift.
The resulting $\Delta\DM$ and $\Delta\FD$ parameters for this simulation are shown in yellow.

The third simulation, ``Scatter Broadening," is the same as ``\DM \& Profile Evolution" but here the frequency-dependent profiles have been convolved with an exponential defined by a single mean scattering time scale, given in Table \ref{tab:pulsar_params}.
As pulse scatter broadening is also a frequency-dependent effect, we expect it to have some small effect on $\widehat{\DMX}$ and $\widehat{\FD}$ \citep{Rickett1977, Levin2016}.
However, since for this simulation only a constant value of $\tau_{\rm{d}}$ is injected, we expect the \FD parameters to account for most, if not all of this variation \citep{Zhu2015, Arzoumanian2016}.
The resulting $\Delta\DM$ and $\Delta\FD$ parameters for this simulation are shown in dark blue.

The final simulation of this set, ``Time-Variable Scatter Broadening", is the same as ``Scatter Broadening" but here we have randomly sampled values of $\tau_{\rm{d}}$ to be injected at each epoch as described in \S \ref{subsec:sim_methods}.
This simulation represents the most realistic of our simulations.
Since here $\tau_{\rm{d}}$ changes, we expect $\widehat{\DMX}$ to be effected more substantially as the \FD parameters are fit over the entire data set, not epoch to epoch.
The resulting $\Delta\DM$ and $\Delta\FD$ parameters for this simulation are shown in magenta.

As with the previous set of simulations, all of the values shown in Figures \ref{fig:PINT_FD_J1744}, \ref{fig:PINT_FD_B1855}, \ref{fig:PINT_FD_B1953}, and \ref{fig:PINT_DMX_Diffs_2} come from fitting for all parameters: \DMX, all \FD parameters, and a JUMP.
We report $\cal{R}_{\rm{rms}}$, $\chi^{2}_{\rm{r}}$, $\Delta\DM_{\rm{rms}}$, $\Delta\FD_{\rm{rms}}$, and the fit value of the JUMP for this set of fit model parameters, as well as the additional model fitting done with a single JUMP, just \DMX and a single JUMP, and just all \FD parameters and a single JUMP, in Tables \ref{tab:J1744-1134_fit_results}, \ref{tab:B1855+09_fit_results}, and \ref{tab:B1953+29_fit_results}.
We note that for this set of simulations, when computing $\Delta\FD_{\rm{rms}}$, two different sets of \FD parameters were used, one set that was fit for using \DMX, \FD, and a JUMP, and one where just \FD and a JUMP was used, both sets come from the ``Profile Evolution" simulation.
This is because the \FD parameters are covariant with \DMX and the values change slightly depending on what parameters are fit for.

\section{Results}
\label{sec:results}

Here we describe the results of the simulations described in \S \ref{sec:sim_data} for each pulsar.
Figures \ref{fig:PINT_DMX_Diffs_1}, \ref{fig:PINT_FD_J1744}, \ref{fig:PINT_FD_B1855}, \ref{fig:PINT_FD_B1953}, and \ref{fig:PINT_DMX_Diffs_2} show the results of simulation analyses when we fit for all parameters (\DMX, all \FD parameters, and a JUMP).
Figures \ref{fig:PINT_DMX_Diffs_1} and \ref{fig:PINT_DMX_Diffs_2} shows the $\Delta\DM$ (the difference between the injected and recovered \DM, $\widehat{\DMX}-\DMX$) for each simulated epoch for all MSPs.
Each set of simulations is split into two sets, Figure \ref{fig:PINT_DMX_Diffs_1} shows the simulations described in \S \ref{subsec:Initial_sims} Figure \ref{fig:PINT_DMX_Diffs_2} shows the $\Delta\DM$ for the simulations described in \S \ref{subsec:Interesting_Sims}.
In these Figures, we again note that the $\Delta\DM$ values have been mean subtracted as described in \S \ref{subsec:Initial_sims}, so we expect all points to be scattered around a mean of zero.
This allows for better visualization of the spread in $\Delta\DM$ between different simulations, where a tighter spread indicates more precise recovery of the injected values.
We also note that there are a few $\Delta\DM$ values that have particularly large error bars.
This is an artifact of the $\widehat{\DMX}$ bin sizes.
Points with these larger uncertainties only have higher frequency 1400~MHz simulated observations within the fifteen or six day window leading to a less accurate $\widehat{\DMX}$.

Figures \ref{fig:PINT_FD_J1744}, \ref{fig:PINT_FD_B1855}, and \ref{fig:PINT_FD_B1953} show the $\Delta\FD$ parameters (the difference between the injected and recovered \FD parameters, $\widehat{\FD_{\rm{i}}}-\FD_{i}$) for for all simulated MSPs, with the top and bottom sets of panels broken up by simulation.
The left hand plots in these Figures show $\Delta\FD$ for each individual \FD parameter in each MSP.
The right hand plots show the total time delay described by the $\Delta\FD$ parameters calculated using Eq. \ref{eq:FD_delay}, as a function of radio frequency.
For all simulations without scatter broadening, when we fit for all parameters, the resulting $\Delta\FD$ parameters are distributed around zero, and within $1\sigma$ of the injected values as shown in Figures \ref{fig:PINT_DMX_Diffs_1}, \ref{fig:PINT_FD_J1744}, \ref{fig:PINT_FD_B1855}, \ref{fig:PINT_FD_B1953}, and  \ref{fig:PINT_DMX_Diffs_2}.

We also compare how fitting for different combinations of \DMX, \FD parameters, and a JUMP affect both the timing residuals, quantified by $\cal{R}_{\rm{rms}}$, and the timing model, quantified by $\chi^{2}_{\rm{r}}$, as reported in Tables \ref{tab:J1744-1134_fit_results}, \ref{tab:B1855+09_fit_results}, and \ref{tab:B1953+29_fit_results}.
These tables also list the values of $\Delta\DM_{\rm{rms}}$ and $\Delta\FD_{\rm{rms}}$, which are used to determine how precisely $\widehat{\DMX}$ and $\widehat{\FD}$ are recovered. 
A large value means that the parameters are recovered less precisely, while smaller values indicate a more precise recovery.

As expected, we see that fitting for additional parameters, e.g., adding \FD parameters even when none have been injected into the simulation, does not negatively impact the $\cal{R}_{\rm{rms}}$ for any simulated data sets.
The $\chi^{2}_{\rm{r}}$ for each fit also appear to be generally unaffected by the addition of more model parameters, however this is due to both the slightly decreased $\chi^{2}$ value of these fits, and the reporting of $\chi^{2}_{\rm{r}}$ to only two decimal places.
Further, when adding additional parameters to the simulations, e.g., \DM variations or \FD parameters, the recovered $\cal{R}_{\rm{rms}}$ when all injected parameters are fit for agree as expected, confirming our methods.

\subsection{Discussion of Frequency-Independent Profile Simulations} \label{subsec:results_initial}

As \FD parameters primarily model variations in the pulse profile with observing frequency \citep{Zhu2015, Arzoumanian2016}, we expect that $\widehat{\FD}$ should all be consistent with zero, and $\Delta\FD_{\rm{rms}}$ should be very small for these simulations .
The exception would be if the profiles are directly shifted in time, or altered in some way (e.g., scatter broadening) as a function of frequency, as denoted in Table \ref{tab:sims_description}.
The injected spectral index does not alter the shape or the profiles, and should not cause additional variations in the FD parameters.
This is indeed what we find, as shown by the upper right panels of Figures \ref{fig:PINT_FD_J1744}, \ref{fig:PINT_FD_B1855}, \ref{fig:PINT_FD_B1953}.

In the ``No Variations" simulations, we find that regardless of what combination of parameters are fit for, we recover almost the same $\cal{R}_{\rm{rms}}$.
This shows that adding additional parameters does not negatively impact the precision of our pulsar timing and confirms that they are not absorbing any additional non-frequency-dependent (or white) noise in the simulated data.

For the ``Time-Variable Scatter Broadening w/ Constant \DM", we find that for all simulated pulsars, the $\cal{R}_{\rm{rms}}$ and $\chi^{2}_{\rm{r}}$ are slightly larger than for the ``No Variation" simulation.
The effect of the scattering delays are obvious from the light green points in Figure \ref{fig:PINT_DMX_Diffs_1}, where the larger the average value, and hence spread of, $\tau_{\rm{d}}$, the less accurate and more variable the resulting $\Delta\DM$, and subsequently $\widehat{\DMX}$, is.
This is less obvious in Figure \ref{fig:PINT_FD_J1744}, but in Figures \ref{fig:PINT_FD_B1855} and \ref{fig:PINT_FD_B1953}, the inability to recover accurate \FD parameters due to larger average injected $\tau_{\rm{d}}$ values is apparent as the light green curve is not consistent with zero.
This indicates that the \DMX and \FD parameters cannot appropriately account for time-variable scattering delays, though the larger variation in the $\Delta\DM$ values suggest that the additional delays from scattering are absorbed by $\widehat{\DMX}$, showing a clear covariance between these two frequency-dependent effects.

For all other simulations in this set, the resulting $\cal{R}_{\rm{rms}}$ and $\chi^{2}_{\rm{r}}$ show that when the appropriate parameters are fit for, all frequency-dependent delays are accounted for, affirming our expectations.
When only \FD parameters are injected and all parameters are fit for, $\Delta\FD_{\rm{rms}}$ increases and $\Delta\DM_{\rm{rms}}$ either remains constant or decreases. 
This is indicative of a small covariance between \DMX and the \FD parameters, and shows that \FD parameters are more susceptible to variations than \DMX is when additional frequency-dependent effects are present and fit.
While this is expected \citep{Zhu2015}, it increases our confidence that when there is very little or no scattering, the \FD parameters are absorbing very little of the dispersive delays.
In these cases, such as PSR~J1744--1134, we can be reasonably confident that the injected \DM is being recovered.

In real pulsar timing data, both \DM variations and additional non-$\nu^{-2}$ frequency-dependent effects are present.
The results of this set of simulations shows that we can accurately recover the full injected delay, when the scattering timescale is very small or zero, giving us confidence in both our methods, and our ability to use this set of simulations as a comparison to our more complex simulations.

\subsection{Discussion of Frequency-Dependent Profile Simulations} \label{subsec:results_complex}

In the ``Profile Evolution" simulations, $\chi^{2}_{\rm{r}}$ is consistent when fitting for just \FD parameters and all parameters, for PSRs~J1744--1134 and B1855+09.
However, for PSR~B1953+29, $\chi^{2}_{\rm{r}}$ is lower when fitting for all parameters compared to just \FD parameters.
As PSR~B1953+29 has a much higher \DM, this suggests that $\widehat{\DMX}$ may absorb more of the delays from profile evolution at higher DMs.
It is possible that this is because at higher DMs, the profile evolution may be primarily dominated by scattering.
Since scattering scales in a similarly frequency-dependent way to \DM, $\widehat{\DMX}$ may absorb the effects of scatter broadening more at higher DMs.
We note, however, that as we have simulated only three MSPs it is difficult to verify this.
To fully explore this relationship would require additional simulations of comparable length exploring not only the scale of DM, but also the size of the DM variations and the number of FD parameters, and as such is beyond the scope of this work.
Additionally, since we recover very similar values of $\cal{R}_{\rm{rms}}$ and $\chi^{2}_{\rm{r}}$ for all MSPs using both methods of fitting, $\widehat{\DMX}$ likely fits out very little of this intrinsic profile evolution, which is expected \citep{Zhu2015, Arzoumanian2016}.

In all other simulations in this set, we find that the best values of $\cal{R}_{\rm{rms}}$ and $\chi^{2}_{\rm{r}}$ occur when we fit for all parameters, which is consistent with the previous simulations discussed in \S \ref{subsec:results_initial}.
For the ``Time-Variable Scatter Broadening" simulation, we note that $\cal{R}_{\rm{rms}}$ is comparable to those obtained in the other simulations in this set, despite Figure \ref{fig:PINT_DMX_Diffs_2} showing that when the average $\tau_{\rm{d}}$ is large, $\widehat{\DMX}$ is much less accurate and more variable.
This is consistent with the results from the ``Time-Variable Scatter Broadening w/ Constant \DM" and suggests that the average scatter broadening is completely fit out by the \FD parameters, while the time variations in the injected $\tau_{\rm{d}}$ are primarily absorbed by the \DMX parameters.
This again shows the clear covariance between \DM and scattering.

For PSR~J1744--1134, which has both the lowest \DM and the most \FD parameters of our simulated pulsars, we find that $\Delta\FD_{\rm{rms}}$ decreases while $\Delta\DM_{\rm{rms}}$ stays comparable when going from a constant to time-varying injected $\tau_{\rm{d}}$.
This is in contrast to both PSRs~B1855+09 and B1953+29, which show an increase in $\Delta\DM_{\rm{rms}}$ but a roughly constant $\Delta\FD_{\rm{rms}}$ when $\tau_{\rm{d}}$ varies with time.
It is difficult to determine if this suggests that pulsars with more \FD parameters and/or smaller $\tau_{\rm{d}}$ are less effected by time-varying $\tau_{\rm{d}}$, or if this is an artifact of pulsars we have chosen to simulate.
In all simulations in this set for PSR~J1744--1134 $\Delta\FD_{\rm{rms}}$ is larger when fitting for all parameters than just \FD parameters, which may similarly suggest that the covariance between \DMX and the \FD parameters is larger for more \FD parameters and/or smaller \DM or \DM variations.
In either case, a comprehensive analysis of this potential relationship would involve exploring a large parameter space, mentioned above, that is beyond the scope of this work.

The difference between the constant and time-varying scatter broadening simulations most clearly shows that while there may be a covariance between \DMX and \FD parameters, it is very small.
As the \FD parameters are fit over the full data set, the differences in $\Delta\FD_{\rm{rms}}$ from a constant to time-varying $\tau_{\rm{d}}$ are much smaller than those of $\Delta\DM_{\rm{rms}}$.
Since the \DMX are fit as a piece-wise function over small timescales, they account for most of the additional time-varying scattering delays.
While this means $\widehat{\DMX}$ may not be as accurate, we can see that it does not seem to have a large effect on $\cal{R}_{\rm{rms}}$, since for both scattering simulations in this set this value is comparable to the smallest $\cal{R}_{\rm{rms}}$ in the baseline ``Profile Evolution" simulation.

\begin{figure}[h]
    \centering
    \includegraphics[width=9cm]{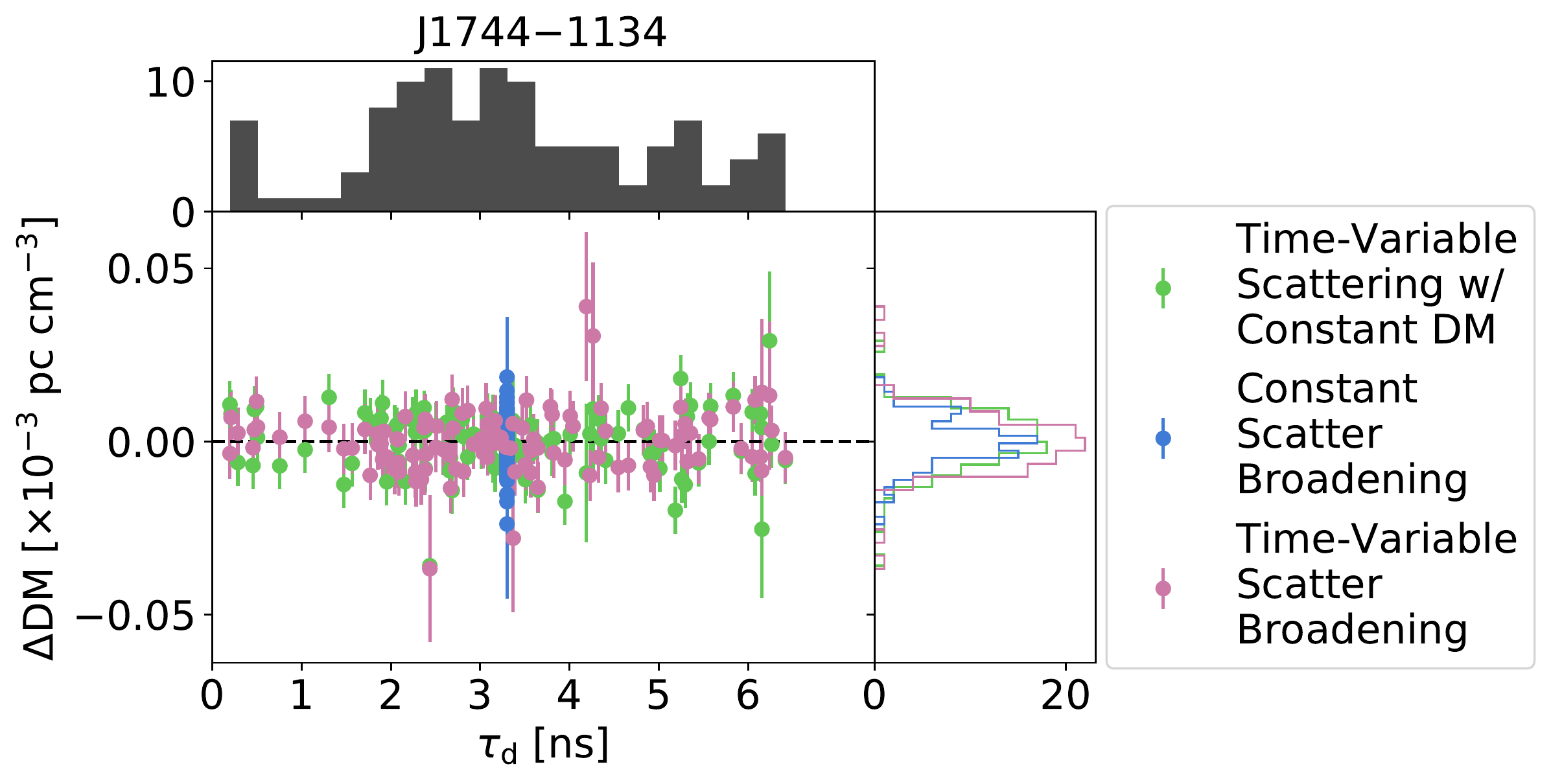}
    \caption{$\Delta\DM$ value ($\widehat{\DMX}-\DMX)$ versus the injected scattering timescale(s) $\tau_{\rm{d}}$ within each \DMX bin for PSR~J1744--1134. There are the same number of points for each simulation shown. We can see that in the case of time-variable scattering delays, the injected \DMX value is recovered less accurately shown by the larger spread in the green and magenta $\Delta\DM$ points compared to the blue points. The top histogram shows the distribution of injected $\tau_{\rm{d}}$ while the right histogram shows the distribution of $\Delta\DM$. We clearly see no correlation between the injected value $\tau_{\rm{d}}$ and the $\Delta\DM$, showing that time-variable scattering only serves to make $\widehat{\DMX}$ more variable and less accurate, though minimally for small average values of  $\tau_{\rm{d}}$.}
    \label{fig:J1744_Scat_DMX}
\end{figure}

\begin{figure}[h]
    \centering
    \includegraphics[width=9cm]{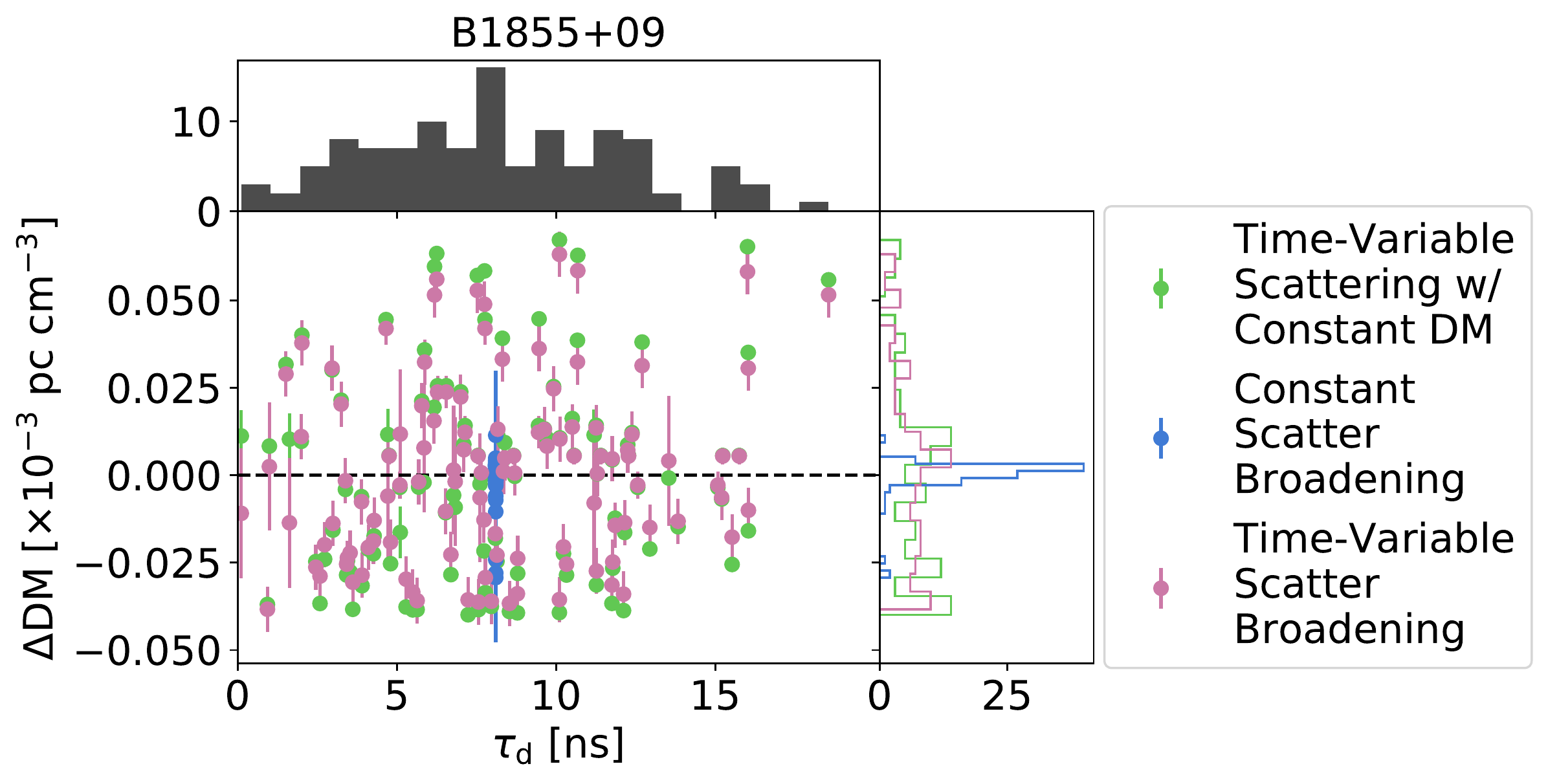}
    \caption{$\Delta\DM$ value ($\widehat{\DMX}-\DMX$) versus the injected scattering timescale(s) $\tau_{\rm{d}}$ within each \DMX bin for PSR~B1855+09. There are the same number of points for each simulation shown. We can see that in the case of time-variable scattering delays, the injected \DMX value is recovered less accurately, as shown by the larger spread in the green and magenta $\Delta\DM$ points compared to the blue points. The top histogram shows the distribution of injected $\tau_{\rm{d}}$ while the right histogram shows the distribution of $\Delta\DM$. We clearly see no correlation between the injected value $\tau_{\rm{d}}$ and the $\Delta\DM$, showing that time-variable scattering only serves to make $\widehat{\DMX}$ more variable and less accurate.}
    \label{fig:B1855+09_Scat_DMX}
\end{figure}

\begin{figure}[h]
    \centering
    \includegraphics[width=9cm]{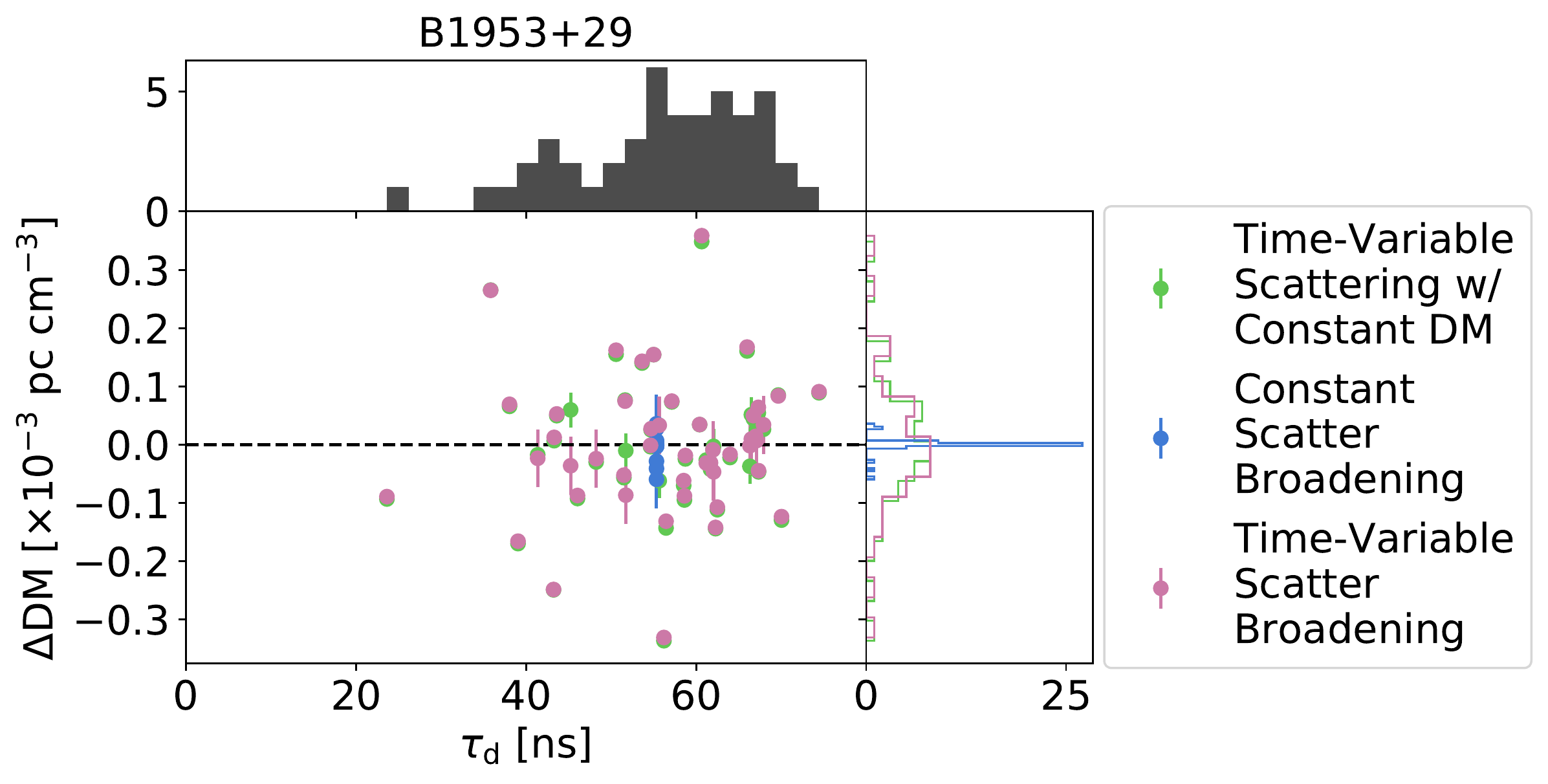}
    \caption{$\Delta\DM$ value ($\widehat{\DMX}-\DMX$) versus the injected scattering timescale(s) $\tau_{\rm{d}}$ within each \DMX bin for PSR~B1953+29. There are the same number of points for each simulation shown. We can see that in the case of time-variable scattering delays, the injected \DMX value is recovered less accurately shown by the larger spread in the green and magenta $\Delta\DM$ points compared to the blue points. The top histogram shows the distribution of injected $\tau_{\rm{d}}$ while the right histogram shows the distribution of $\Delta\DM$. We clearly see no correlation between the injected value $\tau_{\rm{d}}$ and the $\Delta\DM$, showing that time-variable scattering only serves to make $\widehat{\DMX}$ more variable and less accurate.}
    \label{fig:B1953_Scat_DMX}
\end{figure}

\section{Implications for Precision Pulsar Timing}
\label{sec:implications}

The results of our simulations and analyses are important when considering the use of PTAs to detect gravitational waves \citep[e.g.,][]{McLaughlin2013, Kramer2013, Hobbs2013}.
In all simulated MSPs, we find that when profile evolution is present through direct injection or the use of frequency-dependent profiles, $\Delta\FD_{\rm{rms}}$ can change by an order of magnitude when all parameters are fit for, compared to just \FD parameters.
In general this change seems to be an increase for pulsars with more \FD parameters and smaller DMs, and a decrease for pulsars with larger DMs and fewer \FD parameters, however it is important to note that we have only simulated three pulsars in this work.
Regardless, this is evidence of the covariance between \DMX and \FD parameters especially in the simulations where no scattering delays are injected, however the effect on $\cal{R}_{\rm{rms}}$ in negligible, as it is always at a minimum when all appropriate parameters have been fit for.
Additionally, $\chi^{2}_{\rm{r}}$ is almost always closest to one when fitting for all parameters, showing that the addition of \FD parameters does not make the timing model fit worse.

Additionally, one can see from Figures \ref{fig:PINT_DMX_Diffs_1} and \ref{fig:PINT_DMX_Diffs_2} that, when no time-variable scattering delays are injected, as long as the \DMX fit spans both frequency bands, we can recover the injected \DMX, regardless of the \DM variations, the nominal \DM value, or the number of \FD parameters.
This shows that, as expected \citep{Zhu2015, Arzoumanian2016}, our models and fitting are able to, in principle, separate out the physical variations in \DM from any effects modeled out by \FD parameters.
We can conclude that though there is a definitive covariance between \DMX and the \FD parameters, it is very small, and does not affect the precision of the pulsar timing.

The most interesting result and impact found in our simulations is when time-variable $\tau_{\rm{d}}$ are injected.
It is apparent from Figures \ref{fig:PINT_DMX_Diffs_1}, \ref{fig:PINT_FD_J1744}, \ref{fig:PINT_FD_B1855}, \ref{fig:PINT_FD_B1953}, and \ref{fig:PINT_DMX_Diffs_2} that these time-variable delays decrease the accuracy and increase the variability of $\widehat{\DMX}$ as well as $\widehat{\FD}$, and that the level of inaccuracy and variability increases with larger values of $\tau_{\rm{d}}$.
This seems to indicate that the time delays due to scattering are primarily absorbed by the \DMX values indicating a clear covariance between the two frequency-dependent effects.
This shows that in real pulsar data, larger scattering timescales will result in some of the variations in $\widehat{\DMX}$.
However, in our most realistic simulation, ``Time-Variable Scatter Broadening", $\cal{R}_{\rm{rms}}$ is within 10~ns of the minimum expected $\cal{R}_{\rm{rms}}$ given in our baseline ``Profile Evolution" simulation.
So even though the time-variable scattering clearly changes the $\widehat{\DMX}$ and $\widehat{\FD}$, it has a minimal effect on $\cal{R}_{\rm{rms}}$, at least for the three pulsars simulated in this work.

Since $\widehat{\DMX}$ have a much larger spread with varying $\tau_{\rm{d}}$, we also look to see if there is a correlation between the two parameters.
The $\Delta\DM$ values are plotted against the values of $\tau_{\rm{d}}$ injected within each \DMX epoch in Figures \ref{fig:J1744_Scat_DMX}, \ref{fig:B1855+09_Scat_DMX}, and \ref{fig:B1953_Scat_DMX}.
In all cases, we can see that for simulations with time-variable scattering delays (green and magenta), the larger the average value, and hence spread of, $\tau_{\rm{d}}$, the larger the spread in $\Delta\DM$.
For simulations with constant scattering delays (blue), there is very little variability in $\Delta\DM$.
The $\Delta\DM_{\rm{rms}}$ seems to be larger when there are no \DM variations or frequency-dependent profiles injected, though the spread of the resulting $\Delta\DM$ values are roughly the same magnitude between the two simulations.
Most importantly though, there appears to be no correlation between the injected $\tau_{\rm{d}}$ and $\widehat{\DMX}$.
This is likely because the \FD parameters fit out the mean injected $\tau_{\rm{d}}$, so when the distribution of injected $\tau_{\rm{d}}$ is more spread out (see top histograms), $\widehat{\DMX}$ must absorb a larger portion of the scattering delays, resulting in a larger spread in $\Delta\DM$.

The fact that $\cal{R}_{\rm{rms}}$ appears to be relatively unaffected by time-variable scattering delays shows that the delays from frequency-dependent effects are modeled out due to the covariance between \DM and scattering.
As GWs are not radio-frequency dependent, this covariance therefore does not preclude current PTAs from potential detection, though it could induce additional red noise processes in the data.

However, accurate measurements of \DM are extremely important for precision pulsar timing and for understanding noise in the timing data.
In particular, for experiments designed to detect nanohertz GWs, advanced noise modeling techniques such as those discussed in \cite{NG12yrGW2020} will benefit greatly from disentangling the covariances between \DM and scattering where precision down to $\sim100$s of nanoseconds is required \citep[e.g.,][]{Lam2018b,Lam2018d,Lam2019}.
Techniques such as cyclic spectroscopy \citep{Dolch2020, Turner2020}, or alternative methods of quantifying time-variable scattering such as those described in \cite{Main2020} will be necessary to break this covariance.

\section{Conclusions}
\label{sec:conclusions}

Here for the first time we have used the \psrsigsim to simulate pulsar data for three different MSPs with \DM variations, pulse profile evolution with frequency, and time-variable scatter broadening to explore the covariances between these effects.
We show that the \psrsigsim is able to efficiently simulate large amounts of unevenly sampled data spanning long timescales which can then be processed using other standard pulsar software such as \psrchive and \pint to get pulse TOAs as well as fit timing models.
The different delays that were injected into the simulated data such as \DMX, and direct shifts corresponding to the FD parameters can be accurately recovered using these softwares.
This emphasizes not only the usefulness of the \psrsigsim, but also that the standard timing model fitting procedures, such as those implemented in \pint, are able to differentiate between these different frequency-dependent effects.

As an interesting first use case of the \psrsigsim, we explored the covariance between the \DMX and \FD parameters and what, if any, effect this will have on precision pulsar timing.
We find that there is a definite covariance between the two as evidenced by the varying values of $\Delta\FD_{\rm{rms}}$ when fitting for all parameters.
However, in almost all cases, when fitting for all parameters, $\cal{R}_{\rm{rms}}$ was equivalent to the minimum expected values.
This, combined with the fact that the injected values of \DM and \FD parameters were also recovered, shows that this covariance is small, and has a negligible effect on the precision of the pulsar timing. 
While this is expected, these simulations show that this covariance should have little to no impact on the pulsar timing.

Our simulations also find that when scatter broadening is added, the FD parameters are able to fit out the average injected $\tau_{\rm{d}}$.
However, when time-variable scattering delays are injected, both the recovered \DM and \FD parameters, $\widehat{\DMX}$ and $\widehat{\FD}$, are significantly less accurate, increasing with the average injected $\tau_{\rm{d}}$.
We find that most of this additional scattering delay is likely being absorbed by the \DMX parameters, showing the covariance between these effects.
While this does not seem to have a significant impact on $\cal{R}_{\rm{rms}}$, it does imply that some of the variations in \DM seen are due to variable values of $\tau_{\rm{d}}$, and that additional analysis and new techniques, such as cyclic spectroscopy \citep{Dolch2020}, will be needed to separate out these two effects.

These simulations represent just the beginnings of what can be done with the \psrsigsim.
Further studies with the \psrsigsim may look at profile evolution in the era of wide-band pulsar timing \citep{Pennucci2014, NANOGravWB}, or explore other effects not yet incorporated, such as scintillation.
Further improvements to the \psrsigsim to create realistic data sets containing GW signals will be critical for confirming a future detection by PTAs.
Additionally, the \psrsigsim offers ways to explore how future telescope upgrades, such as the ultra-wideband receiver to be installed at the GBT, will affect our pulsar timing \citep{Skipper2019}.
The ability to simulate realistic pulsar data in formats commonly used allow for the potential to test different timing or searching algorithms, explore the effects of different parameters, and test the recovery of input signals, such as GWs, in a new, meaningful way.

\section*{Acknowledgements} \label{acknowledgements}
We would like to thank Paul Baker for useful discussions on \psrsigsim implementations and improvements, Tim Pennucci for useful conversations regarding pulse profile evolution and \FD parameters, and Paul Demorest for discussions regarding \psrfits formatting that were critical for this work.
We would also like to thank the anonymous referee for useful comments and suggestions that contributed to improvements of both this work and the \psrsigsim.
This work was supported by NSF Award OIA-1458952. B.J.S., J.S.H., M.A.M, and M.T.L. are members of the NANOGrav Physics Frontiers Center which is supported by NSF award 1430284.
B.J.S. acknowledges support from West Virginia University through the STEM Mountains of Excellence Fellowship.
The Green Bank Observatory is a facility of the National Science Foundation operated under cooperative agreement by Associated Universities, Inc. 
The Arecibo Observatory is a facility of the National Science Foundation operated under cooperative agreement by the University of Central Florida in alliance with Yang Enterprises, Inc. and Universidad Metropolitana.

\section*{Software} \label{software}
{\textit{Software}: PSRCHIVE \citep{Hotan2004, VanStraten2012}, PyPulse \citep{LamPyPulse}, Matplotlib \citep{Hunter2007}, PINT \citep{Luo2020}, PsrSigSim \citep{Hazboun2020}, SciPy \citep{Jones2001}, NumPy \citep{van2011numpy}, Astropy \citep{astropy:2013, astropy}, pdat \citep{hazboun_pdat}, nanopipe \citep{nanopipe}.}

\startlongtable
\begin{deluxetable*}{cccccc}
\tablecaption{J1744$-$1134 Fitting Results \label{tab:J1744-1134_fit_results}}
\tablecolumns{6}
\tablehead{
\colhead{Fit Parameters} & \colhead{$\cal{R}_{\rm{rms}}$} & \colhead{$\chi^{2}_{\rm{r}}$} & \colhead{$\Delta$DM$_{\rm{rms}}$} & \colhead{$\Delta$FD$_{\rm{rms}}$} & \colhead{JUMP} \\ 
\colhead{} & \colhead{($\upmu$s)} & \colhead{} & \colhead{($10^{-6}$ pc cm$^{-3}$)} & \colhead{($\upmu$s)} & \colhead{($\upmu$s)} 
}
\startdata 
 Jump & $0.14$ & $1.01$ & - & - & $6 \pm 0$ \\ 
 DMX \& Jump & $0.14$ & $1.01$ & $2.4$ & - & $6 \pm 0$ \\ 
 FD \& Jump & $0.14$ & $1.01$ & - & $0.2$ & $6 \pm 0$ \\ 
 DMX \& FD \& Jump & $0.14$ & $1.01$ & $2.4$ & $5.0$ & $6 \pm 0$ \\ 
\hline 
\multicolumn{6}{c}{Simulation: DM Variations} \\ 
\hline 
 Jump & $0.33$ & $10.87$ & - & - & $7 \pm 0$ \\ 
 DMX \& Jump & $0.14$ & $1.03$ & $2.8$ & - & $6 \pm 0$ \\ 
 FD \& Jump & $0.21$ & $4.37$ & - & $2.4$ & $6 \pm 0$ \\ 
 DMX \& FD \& Jump & $0.14$ & $1.03$ & $2.8$ & $14.9$ & $6 \pm 0$ \\ 
\hline 
\multicolumn{6}{c}{Simulation: FD Injection} \\ 
\hline 
 Jump & $33.01$ & $114349.42$ & - & - & $197 \pm 0$ \\ 
 DMX \& Jump & $0.56$ & $11.80$ & $19.1$ & - & $5 \pm 0$ \\ 
 FD \& Jump & $0.14$ & $1.01$ & - & $0.4$ & $6 \pm 0$ \\ 
 DMX \& FD \& Jump & $0.14$ & $1.01$ & $2.7$ & $5.1$ & $6 \pm 0$ \\ 
\hline 
\tablebreak
\multicolumn{6}{c}{Simulation: Time Variable Scattering w/ Constant DM} \\ 
\hline 
 Jump & $0.14$ & $1.02$ & - & - & $6 \pm 0$ \\ 
 DMX \& Jump & $0.14$ & $1.02$ & $34.1$ & - & $6 \pm 0$ \\ 
 FD \& Jump & $0.14$ & $1.02$ & - & $0.7$ & $6 \pm 0$ \\ 
 DMX \& FD \& Jump & $0.14$ & $1.02$ & $34.1$ & $4.6$ & $6 \pm 0$ \\ 
\hline 
\multicolumn{6}{c}{Simulation: DM \& FD Variations} \\ 
\hline 
 Jump & $33.27$ & $116254.67$ & - & - & $198 \pm 0$ \\ 
 DMX \& Jump & $0.56$ & $11.88$ & $17.9$ & - & $5 \pm 0$ \\ 
 FD \& Jump & $0.21$ & $4.34$ & - & $2.3$ & $6 \pm 0$ \\ 
 DMX \& FD \& Jump & $0.14$ & $1.01$ & $2.5$ & $20.0$ & $6 \pm 0$ \\ 
\hline 
\multicolumn{6}{c}{Simulation: Profile Evolution} \\ 
\hline 
 Jump & $1.53$ & $79.40$ & - & - & $7 \pm 0$ \\ 
 DMX \& Jump & $1.19$ & $24.20$ & $9.6$ & - & $1 \pm 0$ \\ 
 FD \& Jump & $1.18$ & $21.42$ & - & - & $2 \pm 0$ \\ 
 DMX \& FD \& Jump & $1.15$ & $21.30$ & $3.8$ & - & $1 \pm 0$ \\ 
\hline 
\multicolumn{6}{c}{Simulation: Profile \& DM Variations} \\ 
\hline 
 Jump & $1.70$ & $112.07$ & - & - & $9 \pm 0$ \\ 
 DMX \& Jump & $1.19$ & $24.09$ & $9.7$ & - & $1 \pm 0$ \\ 
 FD \& Jump & $1.18$ & $22.93$ & - & $2.5$ & $2 \pm 0$ \\ 
 DMX \& FD \& Jump & $1.14$ & $21.19$ & $3.1$ & $7.0$ & $1 \pm 0$ \\ 
\hline 
\multicolumn{6}{c}{Simulation: Constant Scatter Broadening} \\ 
\hline 
 Jump & $1.70$ & $111.96$ & - & - & $9 \pm 0$ \\ 
 DMX \& Jump & $1.19$ & $24.11$ & $10.5$ & - & $1 \pm 0$ \\ 
 FD \& Jump & $1.19$ & $22.95$ & - & $2.3$ & $2 \pm 0$ \\ 
 DMX \& FD \& Jump & $1.15$ & $21.23$ & $3.2$ & $10.8$ & $1 \pm 0$ \\ 
\hline 
\multicolumn{6}{c}{Simulation: Time Variable Scatter Broadening} \\ 
\hline 
 Jump & $1.70$ & $111.94$ & - & - & $9 \pm 0$ \\ 
 DMX \& Jump & $1.19$ & $23.99$ & $10.1$ & - & $1 \pm 0$ \\ 
 FD \& Jump & $1.18$ & $22.82$ & - & $2.4$ & $2 \pm 0$ \\ 
 DMX \& FD \& Jump & $1.14$ & $21.11$ & $3.3$ & $3.5$ & $1 \pm 0$ \\ 
\hline 
\enddata 
\tablecomments{Results of fitting the seven different simulations of PSR~J1744$-$1134 fitting for different parameters, either just a JUMP, $\Delta$DM (DMX) and a JUMP, all FD parameters and JUMP, or $\Delta$DM, all FD parameters, and a JUMP. For each of these four different fits to the seven simulations, we report five quantifiers of the fit. The root mean square of the resulting timing residuals, $\cal{R}_{\rm{rms}}$, where values closer to zero indicate a better fit. The reduced chi-squared of the fit, $\chi^{2}_{\rm{r}}$, where values closer to one indicate a better fit for the number of parameters used in the fit. The root mean square of the $\Delta$DM values, $\Delta$DM$_{\rm{rms}}$, where smaller values indicate that the fit is more accurately recovering the injected values of $\Delta$DM. The root mean square of the $\Delta$FD values, $\Delta$FD$_{\rm{rms}}$, where smaller values indicate that the fit is more accurately recovering the injected FD parameters. We also report the value of the JUMP that is fit in each case.}
\end{deluxetable*}

\startlongtable
\begin{deluxetable*}{cccccc}
\tablecaption{B1855+09 Fitting Results \label{tab:B1855+09_fit_results}}
\tablecolumns{6}
\tablehead{
\colhead{Fit Parameters} & \colhead{$\cal{R}_{\rm{rms}}$} & \colhead{$\chi^{2}_{\rm{r}}$} & \colhead{$\Delta$DM$_{\rm{rms}}$} & \colhead{$\Delta$FD$_{\rm{rms}}$} & \colhead{JUMP} \\ 
\colhead{} & \colhead{($\upmu$s)} & \colhead{} & \colhead{($10^{-6}$ pc cm$^{-3}$)} & \colhead{($\upmu$s)} & \colhead{($\upmu$s)} 
}
\startdata 
 Jump & $0.05$ & $1.49$ & - & - & $120 \pm 0$ \\ 
 DMX \& Jump & $0.05$ & $1.48$ & $1.0$ & - & $120 \pm 0$ \\ 
 FD \& Jump & $0.05$ & $1.49$ & - & $0.1$ & $120 \pm 0$ \\ 
 DMX \& FD \& Jump & $0.05$ & $1.48$ & $1.0$ & $2.3$ & $121 \pm 0$ \\ 
\hline 
\multicolumn{6}{c}{Simulation: DM Variations} \\ 
\hline 
 Jump & $1.55$ & $3652.14$ & - & - & $83 \pm 0$ \\ 
 DMX \& Jump & $0.05$ & $1.53$ & $0.9$ & - & $120 \pm 0$ \\ 
 FD \& Jump & $1.20$ & $2872.38$ & - & $18.2$ & $125 \pm 0$ \\ 
 DMX \& FD \& Jump & $0.05$ & $1.53$ & $0.9$ & $1.6$ & $120 \pm 0$ \\ 
\hline 
\tablebreak
\multicolumn{6}{c}{Simulation: FD Injection} \\ 
\hline 
 Jump & $7.02$ & $37645.77$ & - & - & $-167 \pm 0$ \\ 
 DMX \& Jump & $0.38$ & $118.13$ & $14.0$ & - & $93 \pm 0$ \\ 
 FD \& Jump & $0.05$ & $1.44$ & - & $0.1$ & $120 \pm 0$ \\ 
 DMX \& FD \& Jump & $0.05$ & $1.43$ & $1.3$ & $1.2$ & $121 \pm 0$ \\ 
\hline 
\multicolumn{6}{c}{Simulation: Time Variable Scattering w/ Constant DM} \\ 
\hline 
 Jump & $0.33$ & $109.48$ & - & - & $121 \pm 0$ \\ 
 DMX \& Jump & $0.13$ & $15.43$ & $123.4$ & - & $120 \pm 0$ \\ 
 FD \& Jump & $0.33$ & $108.34$ & - & $1.2$ & $119 \pm 0$ \\ 
 DMX \& FD \& Jump & $0.13$ & $14.57$ & $123.3$ & $2.3$ & $120 \pm 0$ \\ 
\hline 
\multicolumn{6}{c}{Simulation: DM \& FD Variations} \\ 
\hline 
 Jump & $8.09$ & $52073.18$ & - & - & $-204 \pm 0$ \\ 
 DMX \& Jump & $0.38$ & $117.69$ & $13.0$ & - & $93 \pm 0$ \\ 
 FD \& Jump & $1.20$ & $2871.78$ & - & $18.2$ & $125 \pm 0$ \\ 
 DMX \& FD \& Jump & $0.05$ & $1.48$ & $1.4$ & $0.2$ & $121 \pm 0$ \\ 
\hline 
\multicolumn{6}{c}{Simulation: Profile Evolution} \\ 
\hline 
 Jump & $1.90$ & $19.04$ & - & - & $117 \pm 0$ \\ 
 DMX \& Jump & $1.69$ & $14.28$ & $2.4$ & - & $88 \pm 0$ \\ 
 FD \& Jump & $1.47$ & $12.31$ & - & - & $35 \pm 1$ \\ 
 DMX \& FD \& Jump & $1.48$ & $12.43$ & $1.6$ & - & $67 \pm 2$ \\ 
\hline 
\multicolumn{6}{c}{Simulation: Profile \& DM Variations} \\ 
\hline 
 Jump & $2.06$ & $22.67$ & - & - & $79 \pm 0$ \\ 
 DMX \& Jump & $1.69$ & $14.25$ & $2.7$ & - & $88 \pm 0$ \\ 
 FD \& Jump & $1.90$ & $20.36$ & - & $18.2$ & $40 \pm 1$ \\ 
 DMX \& FD \& Jump & $1.48$ & $12.41$ & $1.7$ & $2.0$ & $67 \pm 2$ \\ 
\hline 
\multicolumn{6}{c}{Simulation: Constant Scatter Broadening} \\ 
\hline 
 Jump & $2.08$ & $22.62$ & - & - & $80 \pm 0$ \\ 
 DMX \& Jump & $1.70$ & $14.25$ & $3.2$ & - & $89 \pm 0$ \\ 
 FD \& Jump & $1.90$ & $20.10$ & - & $16.8$ & $38 \pm 1$ \\ 
 DMX \& FD \& Jump & $1.47$ & $12.17$ & $1.6$ & $4.1$ & $66 \pm 2$ \\ 
\hline 
\tablebreak
\multicolumn{6}{c}{Simulation: Time Variable Scatter Broadening} \\ 
\hline 
 Jump & $2.05$ & $22.15$ & - & - & $80 \pm 0$ \\ 
 DMX \& Jump & $1.71$ & $14.34$ & $26.0$ & - & $89 \pm 0$ \\ 
 FD \& Jump & $1.87$ & $19.65$ & - & $16.9$ & $38 \pm 1$ \\ 
 DMX \& FD \& Jump & $1.48$ & $12.28$ & $25.9$ & $3.7$ & $66 \pm 2$ \\ 
\hline 
\enddata 
\tablecomments{Results of fitting the nine different simulations of PSR~B1855+09 fitting for different parameters, either just a JUMP, $\Delta$DM (DMX) and a JUMP, all FD parameters and JUMP, or $\Delta$DM, all FD parameters, and a JUMP. For each of these four different fits to the nine simulations, we report five quantifiers of the fit. The root mean square of the resulting timing residuals, $\cal{R}_{\rm{rms}}$, where values closer to zero indicate a better fit. The reduced chi-squared of the fit, $\chi^{2}_{\rm{r}}$, where values closer to one indicate a better fit for the number of parameters used in the fit. The root mean square of the $\Delta$DM values, $\Delta$DM$_{\rm{rms}}$, where smaller values indicate that the fit is more accurately recovering the injected values of $\Delta$DM. The root mean square of the $\Delta$FD values, $\Delta$FD$_{\rm{rms}}$, where smaller values indicate that the fit is more accurately recovering the injected FD parameters. We also report the value of the JUMP that is fit in each case.}
\end{deluxetable*}

\startlongtable
\begin{deluxetable*}{cccccc}
\tablecaption{B1953+29 Fitting Results \label{tab:B1953+29_fit_results}}
\tablecolumns{6}
\tablehead{
\colhead{Fit Parameters} & \colhead{$\cal{R}_{\rm{rms}}$} & \colhead{$\chi^{2}_{\rm{r}}$} & \colhead{$\Delta$DM$_{\rm{rms}}$} & \colhead{$\Delta$FD$_{\rm{rms}}$} & \colhead{JUMP} \\ 
\colhead{} & \colhead{($\upmu$s)} & \colhead{} & \colhead{($10^{-6}$ pc cm$^{-3}$)} & \colhead{($\upmu$s)} & \colhead{($\upmu$s)} 
}
\startdata 
 Jump & $0.18$ & $1.14$ & - & - & $-455 \pm 0$ \\ 
 DMX \& Jump & $0.18$ & $1.13$ & $3.7$ & - & $-455 \pm 0$ \\ 
 FD \& Jump & $0.18$ & $1.14$ & - & $0.1$ & $-455 \pm 0$ \\ 
 DMX \& FD \& Jump & $0.18$ & $1.13$ & $3.7$ & $0.2$ & $-455 \pm 0$ \\ 
\hline 
\multicolumn{6}{c}{Simulation: DM Variations} \\ 
\hline 
 Jump & $4.20$ & $9668.28$ & - & - & $-411 \pm 0$ \\ 
 DMX \& Jump & $0.18$ & $1.21$ & $6.0$ & - & $-455 \pm 0$ \\ 
 FD \& Jump & $4.05$ & $9493.81$ & - & $35.5$ & $-472 \pm 0$ \\ 
 DMX \& FD \& Jump & $0.18$ & $1.21$ & $6.1$ & $0.1$ & $-455 \pm 0$ \\ 
\hline 
\multicolumn{6}{c}{Simulation: FD Injection} \\ 
\hline 
 Jump & $11.08$ & $4592.29$ & - & - & $-650 \pm 0$ \\ 
 DMX \& Jump & $6.82$ & $1734.07$ & $348.3$ & - & $-477 \pm 0$ \\ 
 FD \& Jump & $0.18$ & $1.20$ & - & $0.1$ & $-455 \pm 0$ \\ 
 DMX \& FD \& Jump & $0.18$ & $1.19$ & $4.8$ & $0.2$ & $-455 \pm 0$ \\ 
\hline 
\tablebreak
\multicolumn{6}{c}{Simulation: Time Variable Scattering w/ Constant DM} \\ 
\hline 
 Jump & $1.08$ & $614.52$ & - & - & $-443 \pm 0$ \\ 
 DMX \& Jump & $0.56$ & $12.98$ & $432.4$ & - & $-463 \pm 0$ \\ 
 FD \& Jump & $1.08$ & $580.55$ & - & $18.0$ & $-472 \pm 0$ \\ 
 DMX \& FD \& Jump & $0.27$ & $3.33$ & $433.3$ & $11.5$ & $-460 \pm 0$ \\ 
\hline 
\multicolumn{6}{c}{Simulation: DM \& FD Variations} \\ 
\hline 
 Jump & $10.78$ & $12816.39$ & - & - & $-606 \pm 0$ \\ 
 DMX \& Jump & $6.83$ & $1739.88$ & $343.5$ & - & $-476 \pm 0$ \\ 
 FD \& Jump & $4.05$ & $9491.51$ & - & $35.5$ & $-472 \pm 0$ \\ 
 DMX \& FD \& Jump & $0.19$ & $1.23$ & $5.9$ & $0.1$ & $-455 \pm 0$ \\ 
\hline 
\multicolumn{6}{c}{Simulation: Profile Evolution} \\ 
\hline 
 Jump & $44.50$ & $4657.17$ & - & - & $-443 \pm 0$ \\ 
 DMX \& Jump & $6.22$ & $127.85$ & $208.6$ & - & $1163 \pm 0$ \\ 
 FD \& Jump & $7.28$ & $136.28$ & - & - & $1205 \pm 1$ \\ 
 DMX \& FD \& Jump & $2.30$ & $20.27$ & $7.6$ & - & $1007 \pm 1$ \\ 
\hline 
\multicolumn{6}{c}{Simulation: Profile \& DM Variations} \\ 
\hline 
 Jump & $43.55$ & $4514.72$ & - & - & $-398 \pm 0$ \\ 
 DMX \& Jump & $6.23$ & $128.29$ & $208.3$ & - & $1163 \pm 0$ \\ 
 FD \& Jump & $8.09$ & $228.31$ & - & $29.8$ & $1197 \pm 1$ \\ 
 DMX \& FD \& Jump & $2.30$ & $20.26$ & $6.6$ & $0.9$ & $1007 \pm 1$ \\ 
\hline 
\multicolumn{6}{c}{Simulation: Constant Scatter Broadening} \\ 
\hline 
 Jump & $43.43$ & $4438.66$ & - & - & $-386 \pm 0$ \\ 
 DMX \& Jump & $6.47$ & $137.58$ & $213.4$ & - & $1166 \pm 0$ \\ 
 FD \& Jump & $7.88$ & $217.56$ & - & $47.5$ & $1179 \pm 1$ \\ 
 DMX \& FD \& Jump & $2.30$ & $20.02$ & $7.2$ & $11.1$ & $1000 \pm 1$ \\ 
\hline 
\multicolumn{6}{c}{Simulation: Time Variable Scatter Broadening} \\ 
\hline 
 Jump & $43.46$ & $4450.10$ & - & - & $-386 \pm 0$ \\ 
 DMX \& Jump & $6.47$ & $137.85$ & $234.3$ & - & $1167 \pm 0$ \\ 
 FD \& Jump & $8.02$ & $231.50$ & - & $48.0$ & $1179 \pm 1$ \\ 
 DMX \& FD \& Jump & $2.31$ & $20.09$ & $99.2$ & $11.5$ & $1001 \pm 1$ \\ 
\hline 
\enddata 
\tablecomments{Results of fitting the nine different simulations of PSR~B1953+29 fitting for different parameters, either just a JUMP, $\Delta$DM (DMX) and a JUMP, all FD parameters and JUMP, or $\Delta$DM, all FD parameters, and a JUMP. For each of these four different fits to the nine simulations, we report five quantifiers of the fit. The root mean square of the resulting timing residuals, $\cal{R}_{\rm{rms}}$, where values closer to zero indicate a better fit. The reduced chi-squared of the fit, $\chi^{2}_{\rm{r}}$, where values closer to one indicate a better fit for the number of parameters used in the fit. The root mean square of the $\Delta$DM values, $\Delta$DM$_{\rm{rms}}$, where smaller values indicate that the fit is more accurately recovering the injected values of $\Delta$DM. The root mean square of the $\Delta$FD values, $\Delta$FD$_{\rm{rms}}$, where smaller values indicate that the fit is more accurately recovering the injected FD parameters. We also report the value of the JUMP that is fit in each case.}
\end{deluxetable*}


\bibliography{SigSim_bib}{}

\end{document}